\documentclass[11pt,a4paper]{article}
\usepackage{graphicx}
\usepackage{float}
\usepackage{afterpage}
\usepackage{epsfig,cite}
\usepackage{amssymb}
\usepackage{amsmath}
\usepackage{dsfont}
\usepackage{multirow}
\usepackage{url,hyperref}
\usepackage{booktabs}
\usepackage{tabularx}
\usepackage{xcolor}
\usepackage{bm}
\usepackage[symbol]{footmisc}
\usepackage{textcomp}
\usepackage{caption}
\usepackage[normalem]{ulem}

\usepackage{url}
\usepackage{hyperref}

\begin{document}

\vspace{-2.0cm}

\begin{center} 

  {\Large \bf A determination of unpolarised pion
    fragmentation functions\\
    \vspace{0.2cm} using semi-inclusive deep-inelastic-scattering
    data: \texttt{MAPFF1.0}\footnote[2]{MAP is an acronym that stands
      for ``Multi-dimensional Analyses of Partonic distributions'' and
      that we adopted as a name for a collaboration of people engaged
      in the study of the three-dimensional structure of hadrons. Visit \href{https://github.com/MapCollaboration}{https://github.com/MapCollaboration} for more information and \href{https://github.com/MapCollaboration/MontBlanc}{https://github.com/MapCollaboration/MontBlanc} for access to the public code used in this analysis.}}
  \vspace{.7cm}

Rabah Abdul Khalek$^{1,2}$, Valerio~Bertone$^{3}$, Emanuele R. Nocera$^4$

\vspace{.3cm}
{\it ~$^1$ Department of Physics and Astronomy, Vrije Universiteit,\\
  Amsterdam, 1081 HV, The Netherlands.}\\
{\it ~$^2$ NIKHEF Theory Group, Science Park 105, 1098 XG Amsterdam,
  The Netherlands.}\\
{\it ~$^3$ IRFU, CEA, Universit\'e Paris-Saclay, F-91191
  Gif-sur-Yvette, France.}\\
{\it ~$^4$ The Higgs Centre for Theoretical Physics, University of Edinburgh,\\
JCMB, KB, Mayfield Rd, Edinburgh EH9 3JZ, Scotland.\\}

\end{center}

\vspace{0.1cm}

\begin{center}
  {\bf \large Abstract}\\
\end{center}
  
We present \texttt{MAPFF1.0}, a determination of unpolarised
charged-pion fragmentation functions (FFs) from a set of
single-inclusive $e^+e^-$ annihilation and lepton-nucleon
semi-inclusive deep-inelastic-scattering (SIDIS) data. FFs are
parametrised in terms of a neural network (NN) and fitted to data
exploiting the knowledge of the analytic derivative of the NN itself
w.r.t.~its free parameters. Uncertainties on the FFs are determined by
means of the Monte Carlo sampling method properly accounting for all
sources of experimental uncertainties, including that of parton
distribution functions. Theoretical predictions for the relevant
observables, as well as evolution effects, are computed to
next-to-leading order (NLO) accuracy in perturbative QCD. We exploit the
flavour sensitivity of the SIDIS measurements delivered by the HERMES
and COMPASS experiments to determine a minimally-biased set of seven
independent FF combinations. Moreover, we discuss the quality of the
fit to the SIDIS data with low virtuality $Q^2$ showing that, as
expected, low-$Q^2$ SIDIS measurements are generally harder to
describe within a NLO-accurate perturbative framework.

\clearpage

\tableofcontents

\section{Introduction}
\label{sec:introduction}

Unpolarised collinear fragmentation functions
(FFs)~\cite{Metz:2016swz} encode the non-perturbative mechanism,
called hadronisation, that leads a fast on-shell parton (a
quark or a gluon) to inclusively turn into a fast hadron moving along
the same direction. In the framework of quantum chromodynamics (QCD),
FFs are a fundamental ingredient to compute the cross section for any
process that involves the measurement of a hadron in the final
state. Among these processes are single-hadron production in
electron-positron annihilation (SIA), semi-inclusive deep-inelastic
scattering (SIDIS), and proton-proton collisions.

An analysis of the measurements for one or more of these processes allows for a
phenomenological determination of FFs. Measurements are compared to the
predictions obtained with a suitable parametrisation of the FFs, which is then
optimised to achieve the best global description possible. The determination
of the FFs has witnessed a remarkable progress in the last years, in particular
of the FFs of the pion, which is the most copiously produced hadron.
Three aspects have been investigated separately:
the variety of measurements analysed, the accuracy of the theoretical settings
used to compute the predictions, and the sophistication of the methodology
used to optimise FFs. In the first respect, global determinations of
FFs including recent measurements for all of the three processes
mentioned above have become available~\cite{deFlorian:2014xna};
in the second respect, determinations of FFs accurate to
next-to-next-to-leading order
(NNLO)~\cite{Anderle:2015lqa,Bertone:2017tyb} or including all-order
resummation~\cite{Anderle:2016czy} have been presented, albeit based
on SIA data only; in the third respect, determinations of FFs using
modern optimisation techniques that minimise parametrisation
bias~\cite{Bertone:2017tyb}, or attempting a simultaneous
determination of the parton distribution functions
(PDFs)~\cite{Moffat:2021dji}, have been performed. These three aspects
have also been investigated for the FFs of the
kaon~\cite{Sato:2016wqj, deFlorian:2017lwf,Bertone:2017tyb,Ethier:2017zbq,
  Sato:2019yez,Moffat:2021dji}.

This paper presents a determination of the FFs of charged pions,
called {\tt MAPFF1.0}, in which the most updated SIA and SIDIS
measurements are analysed to next-to-leading order (NLO) accuracy in
perturbative QCD. Our focus is on a proper statistical treatment of
experimental uncertainties and of their correlations in the
representation of FF uncertainties, and on the efficient minimisation
of model bias in the optimisation of the FF parametrisation. These
goals are achieved by means of a fitting methodology that is inspired
by the framework developed by the NNPDF Collaboration for the
determination of the proton
PDFs~\cite{Ball:2008by,Ball:2012cx,Ball:2014uwa,Ball:2017nwa,Nocera:2014gqa,
  Ball:2013lla}, nuclear
PDFs~\cite{AbdulKhalek:2019mzd,AbdulKhalek:2020yuc}, and
FFs~\cite{Bertone:2017tyb,Bertone:2018ecm}.  The framework combines
the Monte Carlo sampling method to map the probability density
distribution from the space of data to the space of FFs and neural
networks to parametrise the FFs with minimal bias.  In comparison to
previous work~\cite{Bertone:2017tyb,Bertone:2018ecm}, in this paper
the input data set is extended to SIDIS and the neural network is
optimised by means of a gradient descent algorithm that makes use of
the knowledge of the analytic derivatives of the neural network
itself~\cite{AbdulKhalek:2020uza}.

The structure of the paper is as follows. In Sect.~\ref{sec:data} we
introduce the data sets used in this analysis, its features, and the
criteria applied to select the data points. In Sect.~\ref{sec:theory}
we discuss the setup used to compute the theoretical
predictions, focusing on the description of SIDIS multiplicities. In
Sect.~\ref{sec:methodology} we illustrate the methodological framework
adopted in our analysis, specifically the treatment of the
experimental uncertainties and the details of the neural network
parametrisation. In Sect.~\ref{sec:results} we present the results of
our analysis, we assess the interplay between SIA and SIDIS data sets
and the stability of FFs upon variation of the kinematic
cuts. Finally, in Sect.~\ref{sec:conclusion} we provide a summary of
our results and outline possible future developments.

\section{Experimental data}
\label{sec:data}

This analysis is based on a comprehensive set of measurements of
pion-production cross sections in electron-positron SIA and in
lepton-nucleon SIDIS.  In the first case, the data corresponds to the
sum of the cross sections for the production of positively and
negatively charged pions, differential with respect to either the
longitudinal momentum fraction $z$ carried by the fragmenting parton
or the momentum of the measured pion $p_\pi$; the differential cross section is
usually normalised to the total cross section (see Sect.~2.2 in
Ref.~\cite{Bertone:2017tyb} for details). In the second case, the data
corresponds to the hadron multiplicity, that is the SIDIS cross
section normalised to the corresponding inclusive DIS cross section
(see Sect.~\ref{sec:theory} for details). Multiplicities are measured separately
for the production of positively and negatively charged pions.

In the case of SIA, we consider measurements performed at CERN
(ALEPH~\cite{Buskulic:1994ft}, DELPHI~\cite{Abreu:1998vq} and
OPAL~\cite{Akers:1994ez}), DESY
(TASSO~\cite{Brandelik:1980iy,Althoff:1982dh, Braunschweig:1988hv}),
KEK (BELLE~\cite{Leitgab:2013qh} and TOPAZ~\cite{Itoh:1994kb}), and
SLAC (BABAR~\cite{Lees:2013rqd}, TPC~\cite{Aihara:1988su} and
SLD~\cite{Abe:2003iy}). These experiments cover a range of
centre-of-mass energies between $\sqrt{s}\sim 10$~GeV and
$\sqrt{s}= M_Z$, where $M_Z$ is the mass of the $Z$ boson. In the case
of SIDIS, we consider measurements performed at CERN by
COMPASS~\cite{Adolph:2016bga} and at DESY by
HERMES~\cite{Airapetian:2012ki}. The COMPASS experiment utilises a
muon beam with an energy $E_\mu=160$~GeV and a $^6$LiD target. The
HERMES experiment utilises electron and positron beams with an energy
$E_e=27.6$~GeV and hydrogen or deuterium target. Both experiments
measure events within a specific fiducial region. The features of SIA
and SIDIS data are summarised in Tab.~2.1 of
Ref.~\cite{Bertone:2017tyb} and in Tab.~\ref{tab:SIDIS},
respectively. Specific choices that concern some of the available data
sets are discussed below.

\begin{table}[!t]
  \centering
  \small
  \renewcommand{\arraystretch}{1.25}
  \begin{tabularx}{\textwidth}{lcclcX}
    \toprule Data set & Ref.  & $N_{\rm dat}$ & Targets &
    $E_{\rm beam}$ [GeV] & Fiducial cuts ($W\geq W_{\rm low}$;
    $y_{\rm low}\leq
    y\leq y_{\rm up}$)\\
    \midrule COMPASS & \cite{Adolph:2016bga} & 314 (622) & $^6$LiD &
    160 & $W_{\rm low}=5$~GeV, $y_{\rm low} = 0.1$, $y_{\rm up}
    = 0.7$ \\
    HERMES & \cite{Airapetian:2012ki} & \ \ \ 8 (72) & H, $^2$H &
    27.6 & $W_{\rm low}=\sqrt{10}$~GeV, $y_{\rm low} = 0.1$,
    $y_{\rm up} = 0.85$\\
    \bottomrule
  \end{tabularx}
  \caption{\small A summary of the features of the SIDIS data included
    in this analysis. For each of the two data sets we indicate the
    reference, the number of data points after (before) kinematic
    cuts, the target, the beam energy, and the experimental cuts on
    the invariant mass of the final state $W$ and of the inelasticity
    $y$ that define the fiducial region.}
  \label{tab:SIDIS}
\end{table}

Concerning the BELLE experiment, we use the measurement corresponding
to an integrated luminosity
$\mathcal{L}_{\text{int}}=68$~fb$^{-1}$~\cite{Leitgab:2013qh} in spite of the
availability of a more recent measurement based on a larger
luminosity $\mathcal{L}_{\text{int}}=558$~fb$^{-1}$~\cite{Seidl:2020mqc}. Because
of the reduced statistical uncertainties (due to the higher luminosity
of the data sample), the second measurement is significantly more
precise than the first. Therefore, the ability to describe this data
set in a global analysis of FFs crucially depends on the control of
the systematic uncertainties. At present, such a control is
unfortunately lacking. Examples are the unrealistically large
asymmetry of the uncertainties (mainly due to the Pythia tune to
correct for initial-state radiation effects) and the unknown degree of
uncertainty correlation across data points. For these reasons, we were
not able to achieve an acceptable description of the data set of
Ref.~\cite{Seidl:2020mqc}, which we exclude in favour of that of
Ref.~\cite{Leitgab:2013qh}. We multiply all data points by a factor
$1/c$, with $c=1.65$. This is required to correct the data for the
fact that a kinematic cut on radiative photon events was applied to
the data sample instead of unfolding the radiative QED effects, see
Ref.~\cite{Leitgab:2013qh} for details.

Concerning the BABAR experiment, two sets of data are available, based
on {\it prompt} and {\it conventional} yields. The difference between
the two consists in the fact that the latter includes all decay
products with lifetime $\tau$ up to $3\times 10^{-1}$~s, while the
former includes only primary hadrons or decay products from particles
with $\tau\lesssim 10^{-11}$~s.  The conventional cross sections are
about 5-15\% larger than the prompt ones.  Although the conventional
sample was derived by means of an analysis which is closer to that
adopted in other experiments, it turns out to be accommodated in the
global fit worse than its prompt counterpart. We therefore include the
prompt cross section in our baseline fit. The same choice was made in
similar analyses~\cite{deFlorian:2014xna,Sato:2016wqj,Bertone:2017tyb}.

For DELPHI and SLD, in addition to the inclusive measurements, we also
include flavour-tagged measurements, whereby the production of the
observed pion has been reconstructed from the hadronisation of all
light quarks ($u$, $d$, $s$) or of an individual $b$ quark. These
measurements are unfolded from flavour-enriched samples by means of
Monte Carlo simulations and are therefore affected by additional
model uncertainties.  Similar samples for the $c$ quark have been
measured by SLD~\cite{Abe:2003iy}. However,
these are not included because we found it difficult to obtain an
optimal description of them in the fit (see
Sect.~\ref{sec:fitquality}). The OPAL experiment has also measured
completely separated flavour-tagged probabilities for a quark to
hadronise in a jet containing a pion~\cite{Abbiendi:1999ry}.  The interpretation of these measurements is ambiguous in
perturbative QCD beyond leading order, therefore they are not included
in this analysis.

The HERMES multiplicities are presented for various projections of the
fully differential measurement in $P_{h\perp}$, $x$, $z$, and $Q^2$:
these are respectively the transverse component of the hadron momentum
$p_\pi$, the momentum fractions carried by the struck and the
fragmenting parton, and the virtuality of the incoming photon. We use
the projected measurement provided as a function of $Q^2$ and $z$ in
single bins in $x$. We discard the bins with $z<0.2$, which are used
to control the model dependence of the smearing-unfolding procedure,
and with $z>0.8$, which lie in the region where the fractional
contribution from exclusive processes becomes sizeable.

\begin{figure}[!t]
 \begin{center}
   \includegraphics[width=0.75\textwidth]{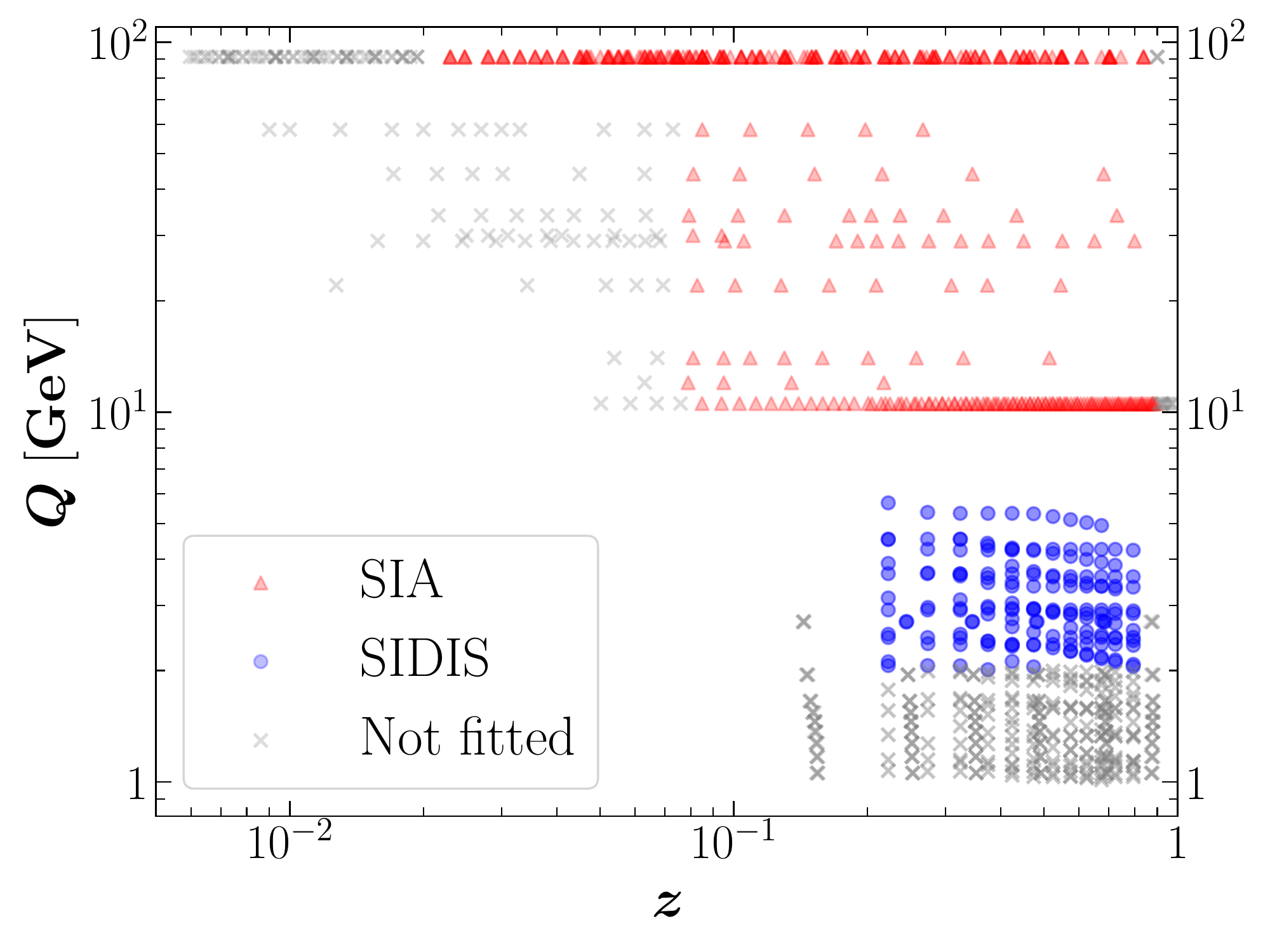}\\
 \end{center}
 \caption{\small Kinematic coverage in the $(z, Q)$ plane of the data
   set included in this analysis. Data points are from SIA (red) and
   SIDIS (blue); gray points are excluded by kinematic cuts.}
 \label{fig:kinplot}
\end{figure}

The kinematic coverage of the data sets included in this analysis is
displayed in Fig.~\ref{fig:kinplot}. As is apparent, SIA and SIDIS data
sets cover two different regions in $Q$: the former range from the
centre-of-mass energy of the $B$-factory measurements, $Q\sim 10$~GeV,
to that of LEP measurements, $Q=M_Z$; the latter, instead, lie at
lower energy scales, $Q\sim 1 - 6$~GeV.
The two data sets are nevertheless complementary. On the one hand,
SIDIS data widens the $Q$ lever arm needed to determine
the gluon FF from perturbative evolution effects. On the other hand,
SIDIS data provides a direct constraint on individual quark and
antiquark FFs, that are otherwise always summed in SIA data.
As expected from kinematic considerations, experiments at higher
centre-of-mass energies provide data at smaller values of $z$.

Kinematic cuts are applied to select only the data points for
which perturbative fixed-order predictions are reliable. For SIA, we
retain the data points with $z$ in the range
$[z_{\rm min}, z_{\rm max}]$; the values of $z_{\rm min}$ and
$z_{\rm max}$ are chosen as in Ref.~\cite{Bertone:2017tyb}:
$z_{\rm min}=0.02$ for experiments at a centre-of-mass energy of $M_Z$
and $z_{\rm min}=0.075$ for all other experiments; $z_{\rm max}=0.9$
for all experiments. For SIDIS, we retain the data points satisfying
$Q>Q_{\rm cut}$, with $Q_{\rm cut}=2$~GeV. This choice maximises the
number of data points included in the fit without spoiling its quality
and follows from a study of the stability of the fit upon choosing
different values of $Q_{\rm cut}$, see Sect.~\ref{sec:Qvar}. 
Overall, after kinematic cuts, we consider $N_{\rm dat}=699$ data
points in our baseline fit almost equally split between SIA
($N_{\rm dat}=377$) and SIDIS ($N_{\rm dat}=322$). Each of the two
processes is dominated by measurements coming respectively from LEP
and $B$-factories, and from COMPASS. 

Information on correlations of experimental uncertainties is taken
into account whenever available. Specifically, for the BABAR
measurement, which is provided with a breakdown of bin-by-bin
correlated systematic uncertainties, for the HERMES measurement,
which is provided with a set of covariance matrices accounting for
correlations of statistical uncertainties obtained from the unfolding
procedure, and for the COMPASS measurement, which is provided with a correlated
systematic uncertainty. Normalisation uncertainties available for the BELLE,
BABAR, TASSO, ALEPH and SLD experiments are assumed to be fully correlated
across all data points in each experiment. If the degree of correlation of
systematic uncertainties is not known, we sum them in quadrature with
the statistical uncertainties. Finally, we symmetrise the systematic uncertainties
reported by the BELLE experiment as described in Ref.~\cite{DAgostini:2004kis}.

\section{Theoretical setup}
\label{sec:theory}

In this section we discuss the theoretical setup used to compute the theoretical
predictions for the SIDIS multiplicities corresponding to the measurements
performed by COMPASS and HERMES. The computation of SIA cross sections and 
the evolution of FFs closely follow Refs.~\cite{Bertone:2017tyb,
  Bertone:2015cwa} and are therefore not discussed here.

We consider the inclusive production of a charged pion, $\pi^\pm$,
in lepton-nucleon scattering:
\begin{equation}
  \label{eq:SIDISreaction}
  \ell(k)+N(p)\rightarrow \ell(k')+ \pi^{\pm}(p_\pi) + X\,.
\end{equation}
The four-momenta involved in this process, along with the definition
$q=k-k'$, allow one to define the following Lorentz invariants that,
under the assumption of a massless target, admit a partonic
interpretation:
\begin{equation}
  \label{eq:kinematics}
\begin{array}{ll}
  \displaystyle Q^2 = -q^2 &\mbox{: (negative) invariant mass of the virtual vector
                             boson,}\\
  \\
  \displaystyle x = \frac{Q^2}{2p\cdot q} & \mbox{: momentum fraction of the nucleon
                                            carried by the incoming parton,}\\
  \\
  \displaystyle z = \frac{p\cdot p_\pi}{p\cdot q}&\mbox{: momentum
                                                 fraction of the outgoing
                                                 parton carried by the pion,}\\
  \\
  \displaystyle y = \frac{Q^2}{xs} & \mbox{: energy transfer (inelasticity)\,,}
\end{array}
\end{equation}
with $\sqrt{s}$ being the centre-of-mass energy of the collision.
Under the assumption $Q\ll M_Z$
(as is the case for all the SIDIS data considered here) only
the exchange of a virtual photon can be considered and the
triple-differential cross section for the reaction in
Eq.~\eqref{eq:SIDISreaction} can be written as
\begin{equation}
  \label{eq:sidis2}
  \frac{d^3\sigma}{dx dQ dz} = 
  \frac{4\pi\alpha^2}{xQ^3} 
  \left[\left(1+(1-y)^2\right) F_2(x,z,Q^2)
    -y^2 F_L(x,z,Q^2) \right]\,,
\end{equation}
where $\alpha$ is the fine-structure constant, and $F_2$ and $F_L$ are
dimensionless structure functions. Within collinear factorisation,
appropriate when $Q\gg \Lambda_{\rm QCD}$, structure functions
factorise as
\begin{equation}\label{eq:f1sidis}
\begin{array}{rcl}
  \displaystyle  F_i(x,z,Q) &=& \displaystyle x\sum_{q\overline{q}} e_q^2 \bigg\{
                                    \left[C_{i,qq}(x,z,Q) \otimes f_q(x,Q) +   
                                    C_{i,qg}(x,z,Q) \otimes f_g(x,Q)\right]\otimes D^{\pi^\pm}_q(z,Q) \\
                                &+&\displaystyle   \left[C_{i,gq}(x,z,Q)
                                    \otimes f_q(x,Q)\right]\otimes
                                    D^{\pi^\pm}_g(z,Q) \bigg\}\,,\qquad i =2,L\,.
\end{array}
\end{equation}
The convolution symbol $\otimes$ acts equally on $x$ and $z$ and has
to be interpreted as follows:
\begin{equation}\label{eq:doubleconvolution}
C(x,z)\otimes f(x)\otimes D(z) =
\int_x^1\frac{dx'}{x'}\int_z^1\frac{dz'}{z'} C(x',z')f\left(\frac{x}{x'}\right) D\left(\frac{z}{z'}\right)\,.
\end{equation}
The sum in Eq.~\eqref{eq:f1sidis} runs over \textit{both} quark and
antiquark flavours active at the scale $Q$, $e_q$ is the electric
charge of the quark flavour $q$, and $f_{q(g)}$ and
$D^{\pi^\pm}_{q(g)}$ denote the collinear quark (gluon) PDFs and FFs,
respectively. Since in this work we are interested in determining the
FFs $D^{\pi^\pm}_{q(g)}$ using existing PDFs $f_{q(g)}$,
Eq.~\eqref{eq:f1sidis} has been arranged in a way that highlights the
role of PDFs as effective charges. Each quark FF
contributing to the cross section is weighted by a factor that, thanks
to PDFs, depends on the specific flavour or antiflavour. As a
consequence, SIDIS cross sections allow for FF quark-flavour separation,
a feature that is not present in SIA cross sections where quark and antiquark
FFs are always summed with equal weight (see {\it e.g.} Eq.~(3.1) in Ref.~\cite{Bertone:2017tyb}).
We use the \texttt{NNPDF31\_nlo\_pch\_as\_0118}~\cite{Ball:2017nwa} as a
reference PDF set. In Sect.~\ref{sec:methodology} we will explain how the
PDF uncertainty is propagated into SIDIS observables and in
Sect.~\ref{sec:theochoices} we will discuss the impact of alternative PDF sets
on the determination of FFs.

The coefficient functions $C$ in Eq.~\eqref{eq:f1sidis} admit the
usual perturbative expansion
\begin{equation}\label{eq:pertexpC}
C(x,z,Q) = \sum_{n=0}\left(\frac{\alpha_s(Q)}{4\pi}\right)^nC^{(n)}(x,z)\,,
\end{equation}
where $\alpha_s$ is the running strong coupling for which we choose
$\alpha_s(M_Z)=0.118$ as a reference value. Presently,
the full set of coefficient functions for both $F_2$ and $F_L$ is only
known up to $\mathcal{O}(\alpha_s)$, \textit{i.e.}
NLO~\cite{Altarelli:1979kv,Graudenz:1994dq}. Explicit $x$- and $z$-space
expressions up to this order can be found for instance in
Ref.~\cite{deFlorian:1997zj} and are implemented in the public code
{\tt APFEL++}~\cite{Bertone:2013vaa, Bertone:2017gds}. A subset of the
$\mathcal{O}(\alpha_s^2)$, \textit{i.e.} NNLO, corrections has been recently
presented in Refs.~\cite{Anderle:2016kwa,Anderle:2018qrw}. However, as long as
the full set of $\mathcal{O}(\alpha_s^2)$ corrections are not known, NNLO
accuracy cannot be attained. For this reason in this analysis
we limit ourselves to NLO accuracy that amounts to considering the
first two terms in the sum in Eq.~\eqref{eq:pertexpC}. For
consistency, also the $\beta$-function and the splitting functions
responsible for the evolution of the strong coupling $\alpha_s$ and of
the FFs, respectively, are computed to NLO accuracy.

So far no heavy-quark mass corrections have been computed for
SIDIS. Therefore, our determination of FFs relies on the so-called
zero-mass variable-flavour-number scheme (ZM-VFNS). In this scheme all
active partons are treated as massless but a partial heavy-quark mass
dependence is introduced by requiring that sub-schemes with different
numbers of active flavours match at the heavy-quark thresholds. Here
we choose $m_c = 1.51$~GeV and $m_b=4.92$~GeV for the charm and bottom
thresholds, respectively, as in the
\texttt{NNPDF31\_nlo\_pch\_as\_0118} PDF set.  In view of the fact
that intrinsic heavy-quark FFs play an important role, it is important
to stress that in our approach inactive-flavour FFs, such as the charm
and bottom FFs below the respective thresholds, are \textit{not} set
to zero. On the contrary, they are allowed to be different from zero
but are kept constant in scale below threshold, \textit{i.e.}  they do
not evolve. This has the consequence that heavy-quark FFs do
contribute to the computation of cross sections also below their
threshold. However, in the specific case of SIDIS this contribution is
suppressed by the PDFs and only appears at NLO.\footnote{This is
  strictly true when using a PDF set that does not include any
  intrinsic heavy-quark contributions as we do here.}

A property of the expressions for the perturbative
coefficient functions $C$ is that the functions $C^{(n)}(x,z)$ with
$n=0,1$ (see Eq.~\eqref{eq:pertexpC}) are bilinear combinations of
single-variable functions:
\begin{equation}
C^{(n)}(x,z) = \sum_t c_tO_t^{(1)}(x)O_t^{(2)}(z)\,,
\end{equation}
where $c_t$ are numerical coefficients. This property enables one to
decouple the double convolution integral in
Eq.~\eqref{eq:doubleconvolution} into a linear combination of single
integrals
\begin{equation}
C^{(n)}(x,z)\otimes f(x)\otimes D(z) = \sum_t
c_t\left[O_t^{(1)}(x)\otimes f(x)\right]\left[O_t^{(2)}(z)\otimes D(z)\right]\,.
\end{equation}
This observation allowed us to considerably speed up the numerical
computation of the SIDIS cross sections.

In order to benchmark the implementation in {\tt APFEL++} used for the
fits and based on the $x$- and $z$-space expressions of
Ref.~\cite{deFlorian:1997zj}, we have carried out a totally
independent implementation of the SIDIS cross section based on the
Mellin-moment expressions~\cite{Stratmann:2001pb,
  Anderle:2012rq}. We made the
Mellin-space version of the cross section publicly available through
the code {\tt MELA}~\cite{Bertone:2015cwa}. The outcome of the
benchmark was totally satisfactory in that in the kinematic region
covered by HERMES and COMPASS the agreement between {\tt APFEL++} and
{\tt MELA} was well below the per-mil level.

The quantity actually measured by both the HERMES and COMPASS
experiments is not an absolute cross section, Eq.~\eqref{eq:sidis2}, but
rather an integrated multiplicity defined as
\begin{equation}\label{eq:multiplicities}
  \frac{dM}{dz} = \left. \left[\int_{Q_{\rm min}}^{Q_{\rm max}}dQ \int_{x_{\rm
          min}}^{x_{\rm max}}dx \int_{z_{\rm min}}^{z_{\rm
          max}}dz\frac{d^3\sigma}{dx dQ dz}\right]\right/
  \left[\Delta z \int_{Q_{\rm min}}^{Q_{\rm max}}dQ \int_{x_{\rm min}}^{x_{\rm max}}dx\frac{d^2\sigma}{dx dQ}\right]\,,
\end{equation}
where the integration bounds define the specific kinematic bin and
$\Delta z = z_{\rm max}-z_{\rm min}$. The denominator is given by the
DIS cross section inclusive w.r.t. the final state that is thus
independent from the FFs. Despite NNLO and heavy-quark mass
corrections are known for the inclusive DIS cross sections, we use the
ZM-VFNS at NLO also in the denominator of
Eq.~\eqref{eq:multiplicities} to match the accuracy of the
numerator. However, we have checked that including NNLO and/or
heavy-quark mass corrections into the inclusive DIS cross section
makes little difference on the determination of FFs.

While the multiplicities measured by the HERMES experiment are binned
in the variables $\{x, Q^2, z\}$, exactly matching the quantity in
Eq.~\eqref{eq:multiplicities}, those measured by the COMPASS
ex\-pe\-ri\-ment are binned in the variables $\{x, y, z\}$ with $y$
defined in Eq.~\eqref{eq:kinematics}. In this case, theoretical
predictions are obtained after adjusting the integration bounds in $Q$
and $x$ in Eq.~\eqref{eq:multiplicities} that become
\begin{equation}
Q_{\rm min} = \sqrt{x_{\rm min}y_{\rm min} s}\,,\qquad Q_{\rm max} = \sqrt{x_{\rm max}y_{\rm max} s}\,,
\end{equation}
and
\begin{equation}\label{eq:compassrepl}
x_{\rm min} \rightarrow {\rm max}\left[x_{\rm min},\frac{Q^2}{sy_{\rm
    max}}\right]\qquad x_{\rm max} \rightarrow {\rm min}\left[x_{\rm max},\frac{Q^2}{sy_{\rm
    min}}\right]\,,
\end{equation}
with $y_{\rm min}$ and $y_{\rm max}$ the bin bounds in $y$. Moreover,
both HERMES and COMPASS measure cross sections within a specific
fiducial region given by
\begin{equation}
W=\sqrt{\frac{(1-x)Q^2}{x}} \geq W_{\rm low}\,,\quad y_{\rm low} \leq
y\leq y_{\rm up}\,,
\end{equation}
with the values of $W_{\rm low}$, $y_{\rm low}$, and $y_{\rm up}$
reported in Tab.~\ref{tab:SIDIS}. These constraints reduce the phase
space of some bins placed at the edge of the fiducial region. The net
effect is that of replacing the $x$ integration bounds in
Eq.~\eqref{eq:multiplicities} with
\begin{equation}
  x_{\rm min}\rightarrow\overline{x}_{\rm min} = {\rm max}\left[x_{\rm min},\frac{Q^2}{sy_{\rm
        up}}\right]\quad\mbox{and}\quad x_{\rm max}\rightarrow\overline{x}_{\rm max} = {\rm min}\left[x_{\rm max},\frac{Q^2}{sy_{\rm
        low}},\frac{Q^2}{Q^2+W_{\rm low}^2}\right]\,,
\end{equation}
with $x_{\rm min}$ and $x_{\rm max}$ to be interpreted as in
Eq.~(\ref{eq:compassrepl}) in the case of COMPASS. We stress that in
our determination of FFs all the integrals in
Eq.~\eqref{eq:multiplicities} are duly computed during the fit. The
effect of computing these integrals, in comparison to evaluating the
cross sections at the central point of the bins, is modest for COMPASS
but sizeable for HERMES.  However, in both cases the integration
contributes to achieving a better description of the data.

Both the HERMES and COMPASS experiments measure multiplicities for
$\pi^+$ and $\pi^-$ separately. However, $\pi^+$ and $\pi^-$ are
related by charge conjugation. In practice, this means that it is
possible to obtain one from the other by exchanging quark and
antiquark distributions and leaving the gluon unchanged:
\begin{equation}
D_{q(\overline{q})}^{\pi^-}(x,Q) =
D_{\overline{q}(q)}^{\pi^+}(x,Q)\,,\quad D_{g}^{\pi^-}(x,Q) =
D_{g}^{\pi^+}(x,Q)\,.
\end{equation}
In this analysis, we use this symmetry to express the $\pi^-$ FFs in
terms of the $\pi^+$ ones and effectively only extract the latter.

We finally note that part of the HERMES measurements and all of the
COMPASS ones are performed on isoscalar targets (deuterium for HERMES
and lithium for COMPASS, see Table~\ref{tab:SIDIS}).
To account for this we have adjusted the PDFs
of the target by using SU(2) isospin symmetry to deduce the neutron
PDFs from the proton ones, which simply amounts to exchanging (anti) up
and (anti) down PDFs, and taking the average between proton and neutron
PDFs. No nuclear corrections are taken into account; no target mass corrections
are considered either, given the complexity to consistently account for them
together with final-state hadron mass corrections~\cite{Guerrero:2015wha}.

\section{Methodology}
\label{sec:methodology}

The statistical framework that we adopted for the inference of the
\texttt{MAPFF1.0} FFs from experimental data relies on the Monte Carlo
sampling method that is nowadays widely used in QCD
analyses~\cite{Ball:2017nwa,
  Nocera:2014gqa,AbdulKhalek:2019mzd,AbdulKhalek:2020yuc,
  Bertone:2017tyb,Bertone:2018ecm
  ,Sato:2016tuz,Sato:2016wqj,Ethier:2017zbq,
  Barry:2018ort,Moutarde:2019tqa,Cuic:2020iwt}.  The main assumption
is that the data originates from a multivariate Gaussian distribution
\begin{equation} \label{eq:gaussian}
\mathcal{G}\left(\bm{x}^{(k)}\right) \propto
\exp{\left[\left(\bm{x}^{(k)} - \bm{\mu}\right)^T \cdot \bm{C}^{-1}\cdot \left(\bm{x}^{(k)} - \bm{\mu}\right)\right]}\, ,
\end{equation}
where
$\bm{x}^{(k)} = \{x_1^{(k)}, x_2^{(k)}, \dddot{},
x^{(k)}_{N_{\text{dat}}}\}$ are equally probable replicas,
$k=1,\dots,N_{\rm rep}$, of a set of $N_{\text{dat}}$ measured quantities.
The expectation value of the distribution corresponding to the measured data
is $\bm{\mu}=\{\mu_1,\,\mu_2,\,\dddot{},\mu_{N_{\text{dat}}}\}$ and
$\bm{C}$ is the covariance matrix that encodes all sources of
uncertainties. The elements of the covariance matrix are defined as follows:
\begin{align} \label{eq:covmat}
  C_{ij}
  =
  \delta_{ij} \sigma_{i,\,\text{unc}}^2  +
  \sum_{\beta} \sigma^{(\beta)}_{i,\,\text{corr}}\sigma^{(\beta)}_{j,\,\text{corr}}\,,
\end{align}
where $\sigma_{i,\,\text{unc}}$ denotes the sum in quadrature of all
uncorrelated uncertainties associated to the $i$-th point and
$\sigma^{(\beta)}_{i,\,\text{corr}}$ is the correlated uncertainty of
source $\beta$ associated to the same point.

The Monte Carlo method aims at propagating the experimental
uncertainties into the space of parameters defined in our case by a
neural network. In order to do so, we generate $N_{\text{rep}}$
replicas of the data, $\bm{x}^{(k)}$, using the Cholesky decomposition
$\bm{L}$ of the covariance matrix $\bm{C}$
($\bm{C} = \bm{L}\cdot \bm{L}^T$) so that
\begin{equation} \label{eq:MCgen}
    \bm{x}^{(k)} = \bm{\mu} + \bm{L}\cdot\bm{r^{(k)}}\,,
\end{equation}
where $\bm{r}^{(k)}$ is an $N_{\text{dat}}$-dimensional normal random
vector such that the full set of replicas satisfies
\begin{equation}
  \frac1{N_{\text{rep}}} {\displaystyle \sum^{N_{\text{rep}}}_k x_i^{(k)}} \simeq \mu_i\, ,\qquad\frac1{N_{\text{rep}}} {\displaystyle \sum^{N_{\text{rep}}}_{k} x_i^{(k)}x_j^{(k)}} \simeq  \mu_i\mu_j + C_{ij}\,,
  \label{eq:MCgeneration}
\end{equation}
in the limit of a sufficiently large number of replicas. We have
verified that the choice $N_{\rm rep}=200$ satisfies
Eq.~\eqref{eq:MCgeneration} with sub-percent accuracy. In the case of
SIDIS, a different proton PDF replica taken at random from the
\texttt{NNPDF31\_nlo\_pch\_as\_0118} set is associated to each replica
$\bm{x}^{(k)}$. This ensures the propagation of PDF uncertainty into the FF uncertainty.

In order to choose the best set of independent FF combinations entering our
parametrisation basis, we study three different cases.
\begin{enumerate}
\item \textbf{11 independent flavours}. This is the most general case
  implied by Eq.~(\ref{eq:f1sidis}) where one aims at disentangling
  all FF flavours and the gluon FF. This parametrisation is overly redundant in
  that the data set used is not able to constrain all 11 combinations.

\item \underline{\textbf{7 independent flavours}}. The sea
  distributions are assumed to be partially symmetric. Specifically,
  $D^{\pi^+}_q=D^{\pi^+}_{\bar{q}}$, for $q=s,c,b$, and
  $D^{\pi^+}_d=D^{\pi^+}_{\bar{u}}$. By doing so, we reduce the number
  of independent distributions down to 7. We observe that under these
  assumptions the quality of the fit does not significantly
  deteriorate with respect to the most general case discussed
  above. In particular, we find it to be the best solution in terms of
  generality and accuracy and therefore we adopt it as our baseline
  parametrisation.

\item \textbf{6 independent flavours}. The approximate SU(2) isospin
  symmetry would suggest that the additional constraint
  $D^{\pi^+}_{u} = D^{\pi^+}_{\bar{d}}$ may hold, further lowering the
  number of independent combinations down to 6. However, it turns out
  that this additional assumption leads to a deterioration
  of the quality of the fit therefore we dropped it.
\end{enumerate}
Finally, the set of 7 independent FF combinations parametrised in our
fit are:
\begin{equation} \label{eq:param_flavours}
\{D^{\pi^+}_u, \quad D^{\pi^+}_{\bar{d}}, \quad D^{\pi^+}_{d}=D^{\pi^+}_{\bar{u}}, \quad D^{\pi^+}_s=D^{\pi^+}_{\bar{s}}, \quad D^{\pi^+}_c=D^{\pi^+}_{\bar{c}}, \quad D^{\pi^+}_b=D^{\pi^+}_{\bar{b}}, \quad D^{\pi^+}_g \}\,.
\end{equation}

The parametrisation is introduced at the initial scale
$\mu_0 = 5\,\text{GeV}$ and consists of a single one-layered
feed-forward neural network $\mathcal{N}_i(z;\bm{\theta})$, where
$\bm{\theta}$ denotes the set of parameters. This network has one
input node corresponding to the momentum fraction $z$, 20 intermediate
nodes with a sigmoid activation function, and 7 output nodes with a
linear activation function corresponding to the flavour combinations
in Eq.~(\ref{eq:param_flavours}). This architecture $[1,\,20,\,7]$
amounts to a total of 187 free parameters. We do not include any
power-like function to control the low- and high-$z$ behaviours,
however we do impose the kinematic constraint $D_i^{\pi^+}(z=1)=0$ by
simply subtracting the neural network itself at $z=1$ as done in
Ref.~\cite{Bertone:2017tyb}. Moreover, we constrain the FFs to be
positive-definite by squaring the outputs. This choice is motivated by
the fact that allowing for negative distributions leads to FFs that
may become unphysically negative. Our parametrisation finally reads
\begin{equation}
    zD^{\pi^+}_i(z, \mu_0 = 5\,\text{GeV}) = \Big(\mathcal{N}_i(z;\bm{\theta}) - \mathcal{N}_i(1;\bm{\theta})\Big)^2\,,
\end{equation}
where the index $i$ runs over the combinations in
Eq.~(\ref{eq:param_flavours}).

The fit is performed by maximising the log-likelihood
$\mathcal{L}\left(\bm{\theta}|\bm{x}^{(k)}\right)$, which is the
probability of observing a given data replica $\bm{x}^{(k)}$ given the
set of parameters $\bm{\theta}^{(k)}$. Together with the multivariate
Gaussian assumption, this is equivalent to minimising the
$\chi^2$. This is defined as
\begin{equation}
\chi^{2(k)} \equiv  \left(\bm{T}(\bm{\theta}^{(k)})- \bm{x}^{(k)}\right)^T \cdot \bm{C}^{-1}\cdot \left(\bm{T}(\bm{\theta}^{(k)})- \bm{x}^{(k)}\right)\,,
\end{equation}
where $\bm{T}(\bm{\theta}^{(k)})$ is the set of theoretical
predictions with the neural-network parametrisation as input. We set
$T_i(\bm{\theta}^{(k)})=\mu_i$ if point $i$ does not satisfy the
kinematic cuts defined in Sect.~\ref{sec:theory} or if it belongs to
the validation set.

We adopt the cross-validation procedure in order to avoid
\textit{overfitting} our FFs. For each data replica, data sets
amounting to more than 10 data points are randomly split into
\textit{training} and \textit{validation} subsets, each containing
half of the points, and only those in the training set are used in the
fit. Data sets with 10 or less data points are instead fully included
in the training set. The $\chi^2$ of the validation set is monitored
during the minimisation of the training $\chi^2$ and the fit is
stopped when the validation $\chi^2$ reaches its absolute minimum.
Replicas whose total $\chi^2$ per point, \textit{i.e.} the $\chi^2$
computed over all data points in the fit, is larger than 3 are
discarded.

The minimisation algorithm adopted for our fit is a
\textit{trust-region} algorithm, specifically the Levenberg-Marquardt
algorithm as implemented in the \texttt{Ceres Solver}
code~\cite{ceres-solver}. This is an open source \texttt{C++} library
for modelling and solving large optimisation problems. The
neural-network parametrisation and its analytical derivatives with
respect to the free parameters are provided by
\texttt{NNAD}~\cite{AbdulKhalek:2020uza}, an open source \texttt{C++}
library that provides a fast implementation of an arbitrarily large
feed-forward neural network and its analytical derivatives.

\section{The {\tt MAPFF1.0} set}
\label{sec:results}

In this section we present the main results of this analysis. In
Sect.~\ref{sec:fitquality} we discuss the quality of our baseline fit
that we dub {\tt MAPFF1.0}. In Sect.~\ref{sec:FFs} we illustrate its
features: we compare our FFs to other recent determinations and we
study the stability of the fit upon the choice of input PDFs and of
the parametrisation scale $\mu_0$. In Sect.~\ref{sec:dataimpact} we
study the impact of some specific data sets: we discuss the origin of
the difficulty in fitting SIA charm-tagged data, we investigate the
impact of the SIDIS data on FFs, and finally we study the dependence
of the fit quality on the low-$Q$ cut on the SIDIS data.

\subsection{Fit quality}
\label{sec:fitquality}

Tab.~\ref{tab:chi2s} reports the value of the $\chi^2$ per data point for
the individual data sets included in the {\tt MAPFF1.0} fit along with the
number of data points $N_{\rm dat}$ that pass the kinematic cuts discussed
in Sect.~\ref{sec:data}. The table also reports the partial $\chi^2$ values
of the SIDIS and SIA data sets separately as well as the global one.

\begin{table}[!t]
\renewcommand{\arraystretch}{1.25}
\centering
\scriptsize
\begin{tabular}{lccccccccc}
  \toprule
  Experiment & $\chi^2$ per point & $N_{\rm dat}$ after cuts \\
  \midrule
  HERMES $\pi^-$ deuteron & 0.60 &  2 \\
  HERMES $\pi^-$ proton & 0.02 &  2 \\
  HERMES $\pi^+$ deuteron & 0.30 &  2 \\
  HERMES $\pi^+$ proton & 0.53 &  2 \\
  COMPASS $\pi^-$ & 0.80 &  157 \\
  COMPASS $\pi^+$ & 1.07 &  157 \\
  \midrule
  Total SIDIS & 0.78 & 322 \\
  \midrule
  BELLE $\pi^\pm$ & 0.09 &  70 \\
  BABAR prompt $\pi^\pm$ & 0.90 &  39 \\
  TASSO 12 GeV $\pi^\pm$ & 0.97 &  4 \\
  TASSO 14 GeV $\pi^\pm$ & 1.39 &  9 \\
  TASSO 22 GeV $\pi^\pm$ & 1.85 &  8 \\
  TPC $\pi^\pm$ & 0.22 &  13 \\
  TASSO 30 GeV $\pi^\pm$ & 0.34 &  2 \\
  TASSO 34 GeV $\pi^\pm$ & 1.20 &  9 \\
  TASSO 44 GeV $\pi^\pm$ & 1.20 &  6 \\
  TOPAZ $\pi^\pm$ & 0.28 &  5 \\
  ALEPH $\pi^\pm$ & 1.29 &  23 \\
  DELPHI total $\pi^\pm$ & 1.29 &  21 \\
  DELPHI $uds$ $\pi^\pm$ & 2.84 &  21 \\
  DELPHI bottom $\pi^\pm$ & 1.67 &  21 \\
  OPAL $\pi^\pm$ & 1.72 &  24 \\
  SLD total $\pi^\pm$ & 1.14 &  34 \\
  SLD $uds$ $\pi^\pm$ & 2.05 &  34 \\
  SLD bottom $\pi^\pm$ & 0.55 &  34 \\
  \midrule
  Total SIA & 1.10 & 377 \\
  \midrule
\textbf{Global data set} & \textbf{0.90} & \textbf{699} \\
  \bottomrule
\end{tabular}
\caption{\small Values of the $\chi^2$ per data point for the
  individual data sets included in the {\tt MAPFF1.0} analysis. The
  number of data points $N_{\rm dat}$ that pass kinematic cuts and the
  SIDIS, SIA, and global $\chi^2$ values are also displayed.}
\label{tab:chi2s}
\end{table}

The global $\chi^2$ per data point, equal to 0.90, indicates a general
very good description of the entire data set. A comparable fit quality
is observed for both the SIDIS and SIA sets separately, with
collective $\chi^2$ values equal to 0.78 and 1.10, respectively. A
closer inspection of Tab.~\ref{tab:chi2s} reveals that an acceptable
description is achieved for all of the individual data sets. Some
particularly small $\chi^2$ values are also obtained. This is the case
of the HERMES $\pi^-$ and BELLE data. In the case of HERMES, the
smallness of the $\chi^2$ is not statistically significant given that
only two data points survive the cuts. The smallness of the $\chi^2$
of BELLE is instead well-known and follows from an overestimate of the
systematic uncertainties~\cite{deFlorian:2014xna, Hirai:2016loo,
  Bertone:2017tyb, Gamberg:2021lgx}.

\begin{figure}[!t]
  \begin{center}
  \includegraphics[width=0.49\textwidth]{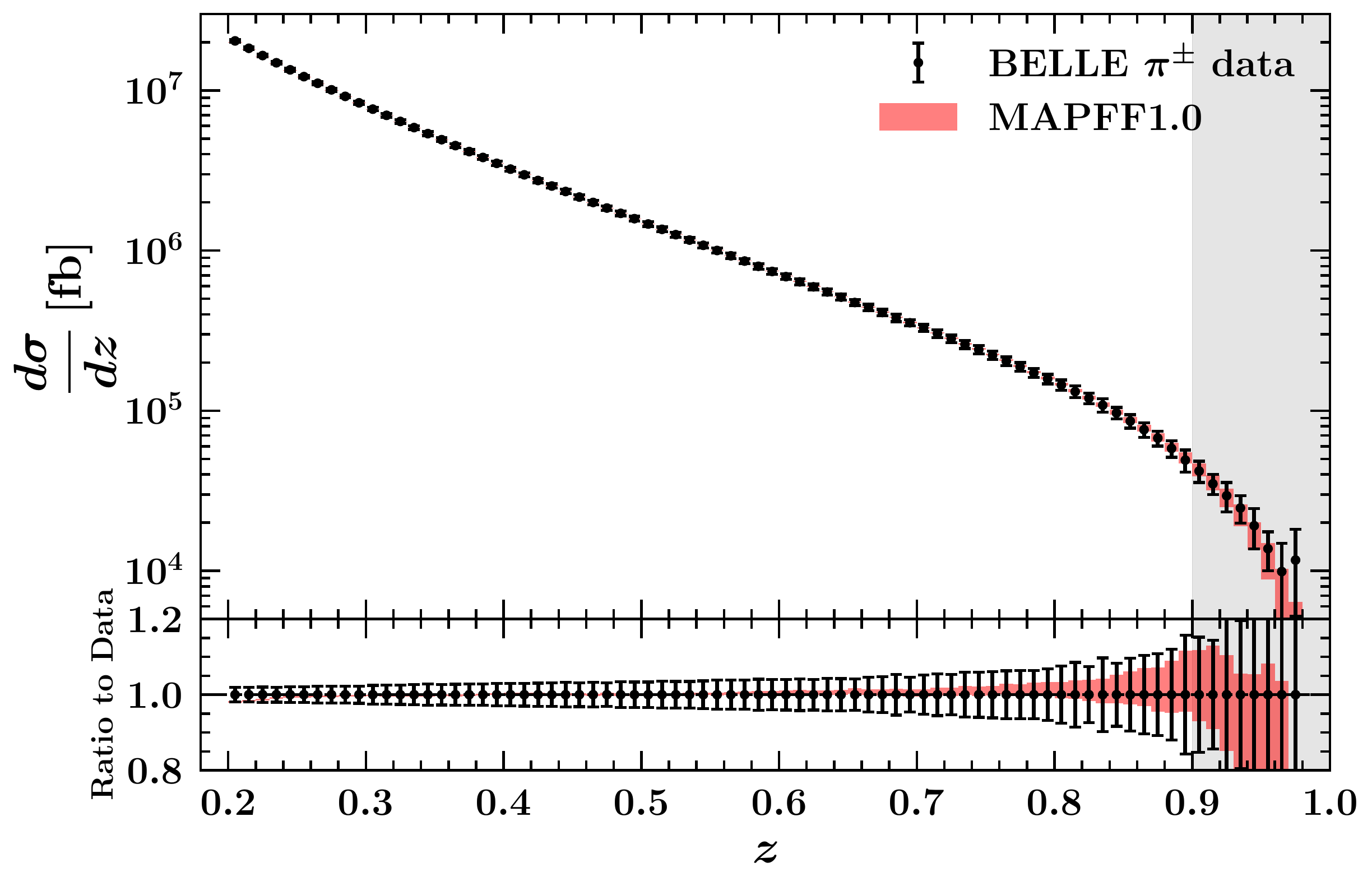}
  \includegraphics[width=0.49\textwidth]{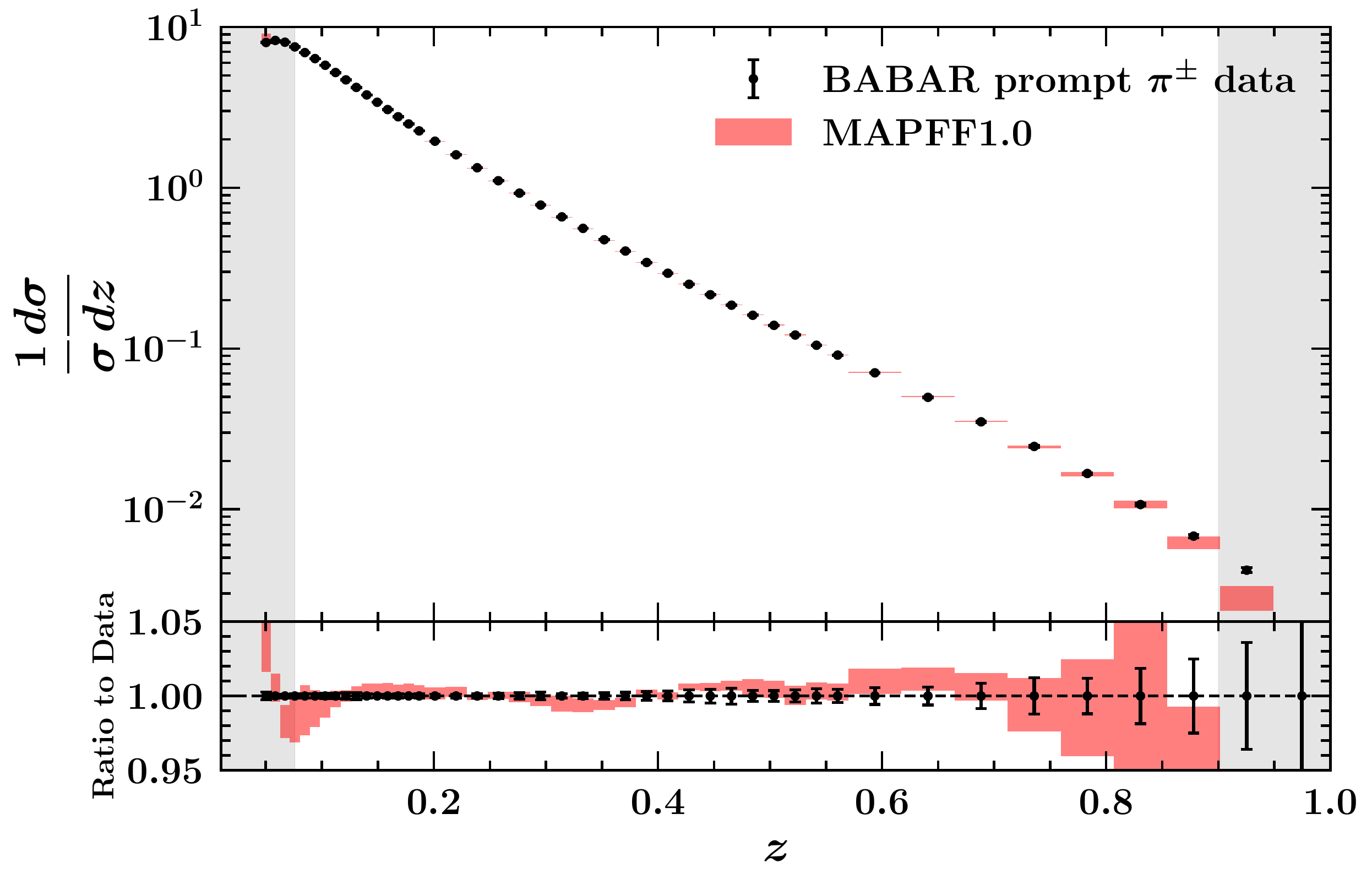}\\
  \includegraphics[width=0.49\textwidth]{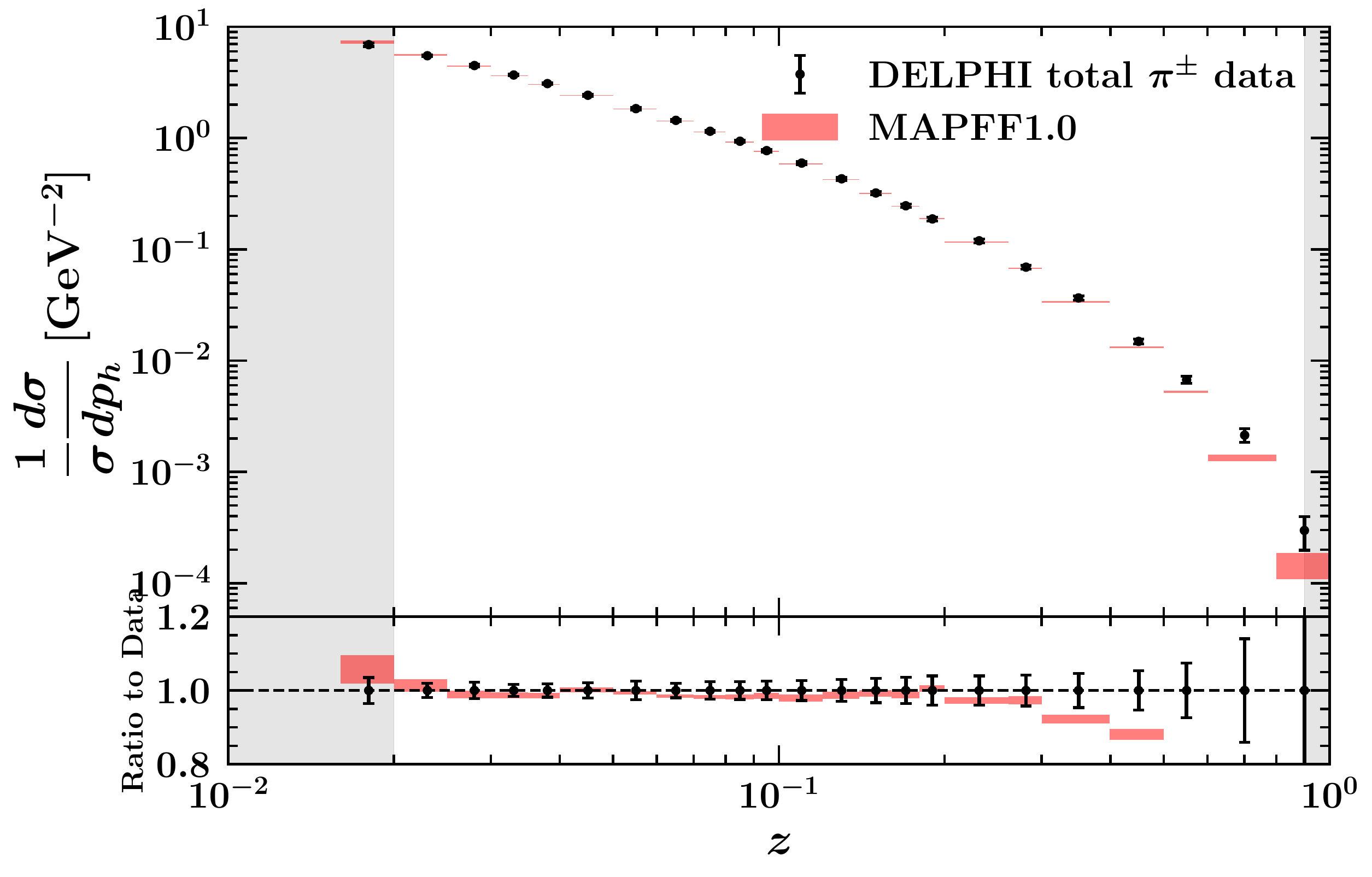}
  \includegraphics[width=0.49\textwidth]{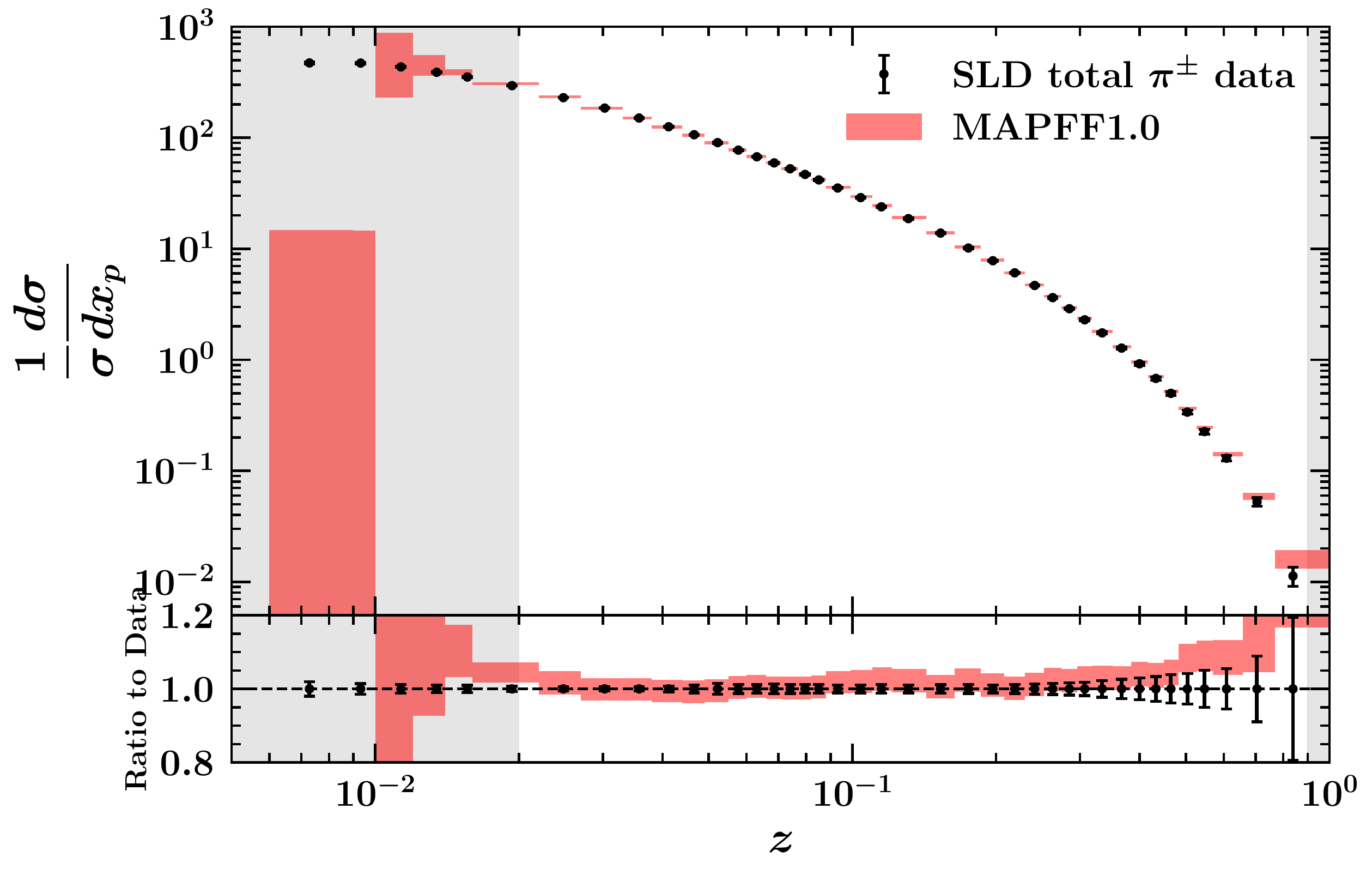}\\
  \end{center}
  \caption{\small Top row: data-theory comparison for the $B$-factory
    experiments BELLE (left) and BABAR (right) at
    $\sqrt{s}\simeq10.5$~GeV.  Bottom row: representative data-theory
    comparison for SIA data sets at $\sqrt{s} = M_Z$ from DELPHI
    (left) and SLD (right). The shaded regions are excluded from the
    fit.}
  \label{fig:BfactoriesMZexps}
\end{figure}

It is instructive to look at the comparison between data and
predictions obtained with the {\tt MAPFF1.0} FFs for some selected
data sets. The top row of Fig.~\ref{fig:BfactoriesMZexps} shows the
comparison for the $B$-factory experiments BELLE and BABAR at
$\sqrt{s}\simeq 10.5$ GeV, while the bottom row shows the comparison
for two representative data sets at $\sqrt{s}=M_Z$, \textit{i.e.} the
total cross sections from DELPHI and SLD. The upper panels display the
absolute distributions while the lower ones the ratio to the
experimental central values. The shaded areas indicate the regions
excluded from the fit by the kinematic cuts. In order to facilitate
the visual comparison, predictions are shifted to account for
correlated systematic uncertainties~\cite{Pumplin:2002vw}, when
present.  Consistently with the $\chi^2$ values reported in
Tab.~\ref{tab:chi2s}, the description of these data sets is very good
within cuts.

\begin{figure}[!t]
  \begin{center}
    \includegraphics[width=\textwidth]{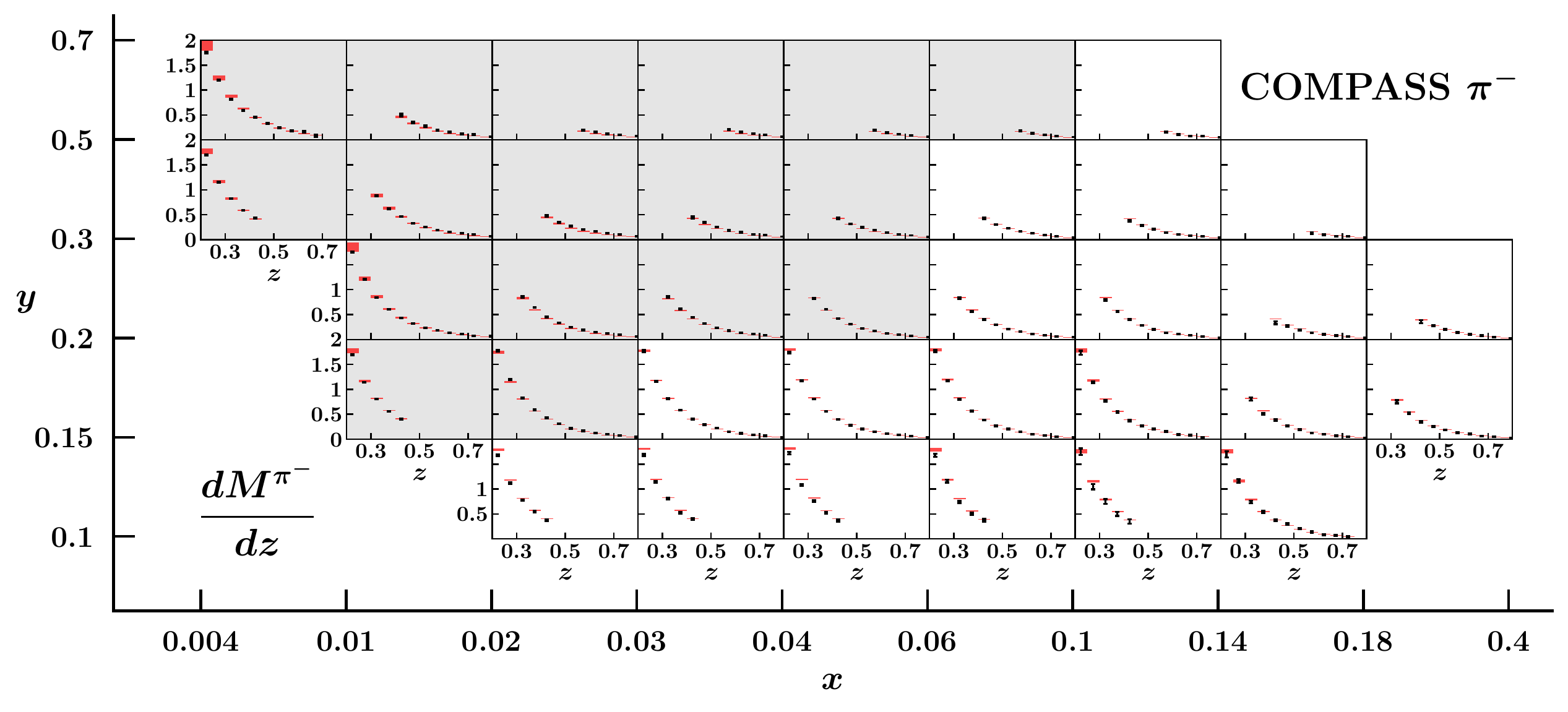}\\
  \end{center}
  \caption{\small Data-theory comparison for the $\pi^-$
    multiplicities from COMPASS. The shaded bins are excluded from the
    fit due to the cut in $Q$.}
  \label{fig:COMPASSall}
\end{figure}

Fig.~\ref{fig:COMPASSall} shows the data-theory comparison for the
COMPASS $\pi^-$ multiplicities. Each panel
displays a distribution in $z$ corresponding to a bin in $x$ and
$y$. As above, theoretical predictions have been shifted to ease the
visual comparison. The grey-shaded panels are not fitted because they
do not fulfil the cut in $Q$ discussed in Sect.~\ref{sec:data}. Once
again, the goodness of the $\chi^2$ values in Tab.~\ref{tab:chi2s} is
reflected in a general very good description of the data. The analogous
plot for $\pi^+$ multiplicities looks qualitatively similar to
Fig.~\ref{fig:COMPASSall}, therefore it is not shown.

\subsection{Fragmentation Functions}
\label{sec:FFs}

We now present our FFs. We first compare them with other FF sets, then
we study the impact of some relevant theoretical choices.

\subsubsection{Comparison with other FF sets}
\label{subsec:otherFFs}

\begin{figure}[!t]
  \begin{center}
    \includegraphics[width=0.9\textwidth]{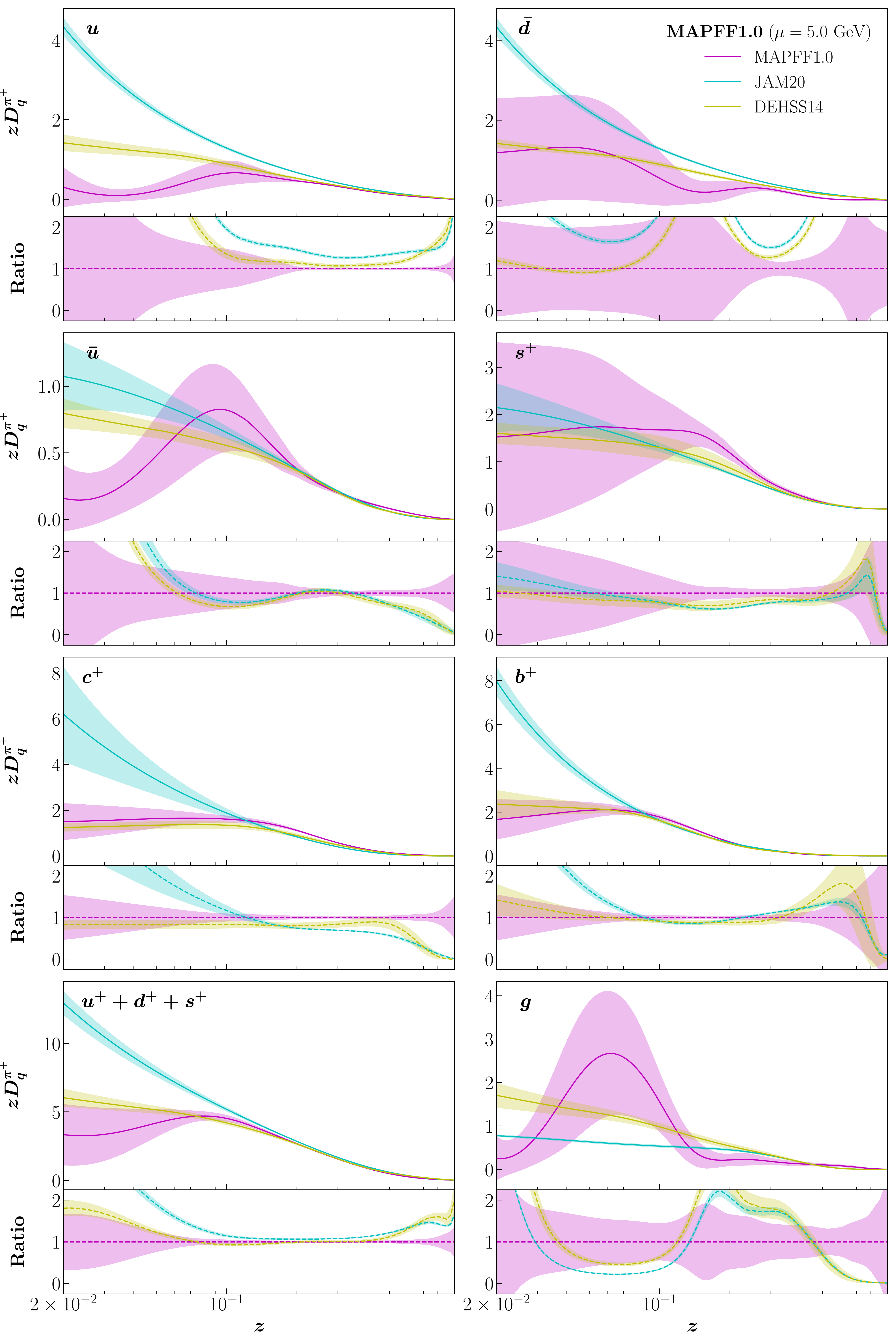}\\
    \end{center}
    \caption{\small Comparison of the \texttt{MAPFF1.0},
      \texttt{DEHSS14}~\cite{deFlorian:2014xna}, and
      \texttt{JAM20}~\cite{Moffat:2021dji} FFs. We display the
      $D_u^{\pi^+}$, $D_{\bar d}^{\pi^+}$, $D_{\bar u}^{\pi^+}$,
      $D_{s^+}^{\pi^+}$, $D_{c^+}^{\pi^+}$, $D_{b^+}^{\pi^+}$,
      $D_{u^++d^++s^+}^{\pi^+}$ and $D_g^{\pi^+}$ FFs at
      $\mu=5$~GeV. For each FF we plot the absolute distributions in
      the upper panel and their ratio to the central value of the
      \texttt{MAPFF1.0} set in the lower one.}
  \label{fig:Baseline}
\end{figure}

In Fig.~\ref{fig:Baseline} we compare the $\pi^+$ FFs obtained from
our baseline fit, \texttt{MAPFF1.0}, to those from
\texttt{JAM20}~\cite{Moffat:2021dji} and
\texttt{DEHSS14}~\cite{deFlorian:2014xna}. All three sets include a
similar SIA and SIDIS data set. However, the \texttt{JAM20} set also
includes inclusive deep-inelastic scattering and fixed-target
Drell-Yan measurements that are used to simultaneously determine PDFs,
while the \texttt{DEHSS14} set also includes pion production
measurements in proton-proton collisions.

In the case of $D_u^{\pi^+}$ and $D_{\bar d}^{\pi^+}$, as is clear
from the upper row of Fig.~\ref{fig:Baseline}, \texttt{JAM20} assumes
SU(2) isospin symmetry which results in
$D_u^{\pi^+} = D_{\bar{d}}^{\pi^+}$. \texttt{DEHSS14} instead assumes
$D_u^{\pi^+} + D_{\bar{u}}^{\pi^+}\propto D_{{d}}^{\pi^+} +
D_{\bar{d}}^{\pi^+}$ where the proportionality factor is a
$z$-independent constant that parameterises any possible isospin
symmetry violation. As explained in Sect.~\ref{sec:methodology}, in
\texttt{MAPFF1.0} we parametrise $D_u^{\pi^+}$ and
$D_{\bar{d}}^{\pi^+}$ independently, thus allowing for a $z$-dependent
isospin symmetry violation which however turned out not to be
significant. For $z\lesssim 0.1$, where experimental data is sparse
(see Fig.~\ref{fig:kinplot}), the relative uncertainty on the
$D_u^{\pi^+}$ and $D_{\bar{d}}^{\pi^+}$ distributions from
\texttt{MAPFF1.0} is larger than that of the corresponding
distributions from \texttt{DEHSS14} and \texttt{JAM20}.  At large $z$,
we observe a suppression of the \texttt{MAPFF1.0} FFs
w.r.t. \texttt{DEHSS14} and \texttt{JAM20} for both $D_u^{\pi^+}$ and
$D_{\bar{d}}^{\pi^+}$. This suppression is compensated by an
enhancement of the sea quarks as we will further discuss below.

In the case of the sea FFs, we find a
good agreement at very large $z$ with both \texttt{DEHSS14} and
\texttt{JAM20} for $D_{s^+}^{\pi^+}$ and $D_{b^+}^{\pi^+}$. At low
$z$, on the one hand, we observe that the sea FFs from
\texttt{DEHSS14}, except $D_{\bar{u}}^{\pi^+}$, and the
$D_{s^+}^{\pi^+}$ FF from \texttt{JAM20} are within the
\texttt{MAPFF1.0} uncertainties. On the other hand, \texttt{JAM20} and
\texttt{DEHSS14} show an enhancement for the rest of the FFs
w.r.t.~\texttt{MAPFF1.0}.

We note that a very good agreement at intermediate $z$ is found for
the light singlet $D_{u^+}^{\pi^+}+D_{d^+}^{\pi^+}+D_{s^+}^{\pi^+}$
combination. This particular combination is most sensitive to
inclusive SIA data and the agreement reflects the fact that all three
collaborations are able to describe this data comparably well. The
gluon FF of \texttt{MAPFF1.0} is affected by large uncertainties. This
is a consequence of using observables that are not directly sensitive
to this distribution. The \texttt{MAPFF1.0}, \texttt{JAM20}, and
\texttt{DEHSS14} gluon FFs remain fairly compatible within
uncertainties.

Finally, we observe that most of the distributions of the
\texttt{MAPFF1.0} set present a turn-over in the region $ 0.1 \lesssim
z \lesssim 0.2$ that is absent in the other two sets. This feature can
be ascribed to the fact that \texttt{MAPFF1.0} implements cuts on the
minimum value of $z$ that are generally lower than those used by the
other two collaborations (see Ref.~\cite{Bertone:2017tyb} for a
detailed study).

\subsubsection{Impact of theoretical choices}
\label{sec:theochoices}

We have studied the stability of our FFs upon the input PDF set used
to compute SIDIS multiplicities. 
In order to assess the impact
of the PDF uncertainty, we performed an additional fit using the
central PDF member of the \texttt{NNPDF31\_nlo\_pch\_as\_0118} set
for all Monte Carlo replicas. We found that neglecting the PDF
uncertainty has a very small impact on the FFs. In addition, we
studied the dependence of the FFs on the specific PDF set by
performing two additional fits using the central member of the
\texttt{CT18NLO}~\cite{Hou:2019efy} and
\texttt{MSHT20nlo\_as118}~\cite{Bailey:2020ooq} sets. Also in this
case, we found that the difference at the level of FFs was very mild.
In fact, a
reduced sensitivity to the treatment of PDFs was to be expected
because our FFs depend on them only through the SIDIS measurements
that are delivered as multiplicities for both by COMPASS and
HERMES. For this particular observable, see
Eq.~\eqref{eq:multiplicities}, PDFs enter both the numerator and the
denominator, hence the sensitivity of the observable to the PDFs
largely cancels out.

We have also studied the stability of our FFs upon the choice of the
parametrisation scale $\mu_0$. To this purpose, we have repeated our
baseline fit by lowering the value of $\mu_0$ from 5~GeV to 1~GeV.
This led to almost identical FFs. We stress that the possibility to
freely choose the parameterisation scale, no matter whether above or
below the heavy quark thresholds, is due to the fact that we do not
set inactive-flavour FFs to zero below their respective threshold (see
Sect.~\ref{sec:theory}).

\subsection{Impact of the data}
\label{sec:dataimpact}

We now justify our exclusion of the SIA charm-tagged data from the fit
as well as our choice of the cut on the virtuality $Q^2$ for the SIDIS
data. We also discuss the separate impact of COMPASS and HERMES on the
FFs.

\subsubsection{Data compatibility}
\label{sec:charmtag}

\begin{figure}[!t]
  \begin{center}
    \includegraphics[width=\textwidth]{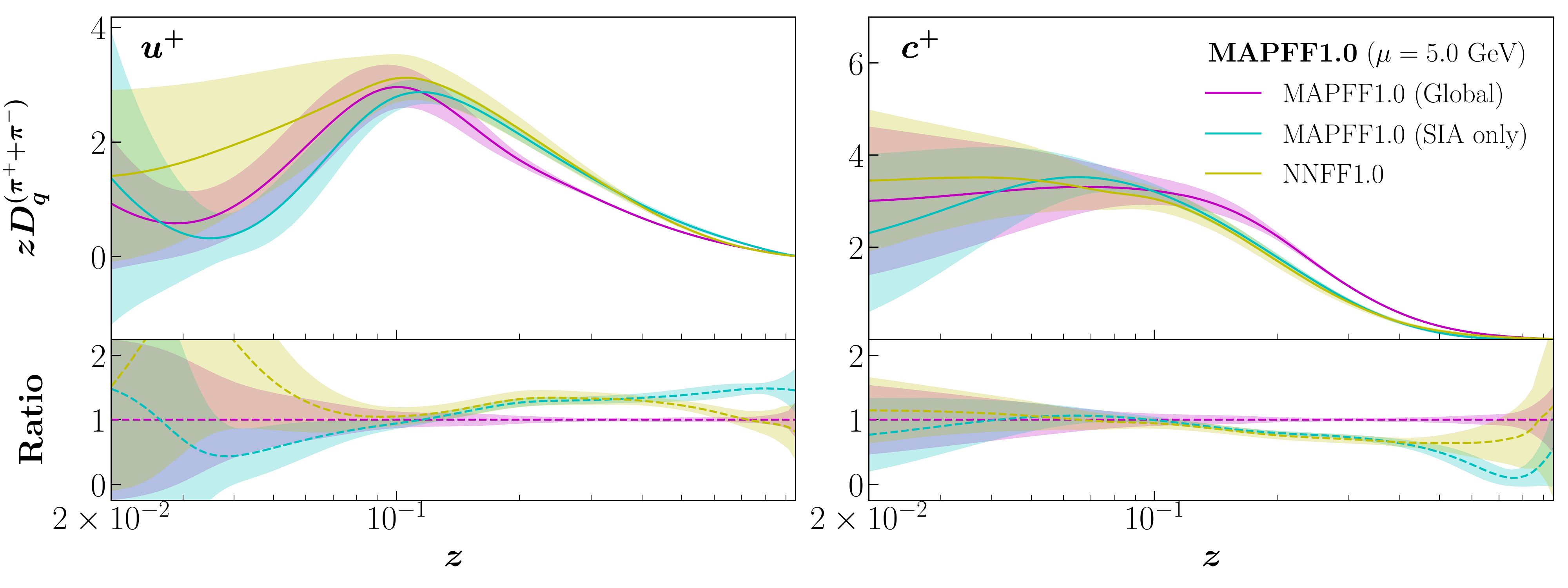}
  \end{center}
  \caption{\small Comparison between the baseline \texttt{MAPFF1.0}
    fit (which includes SIA and SIDIS data), a variant of the {\tt
      MAPFF1.0} fit to SIA data only, and
    \texttt{NNFF1.0}~\cite{Bertone:2017tyb} (which includes only SIA
    data). The left (right) plot shows the $D_{u^{+}}^{\pi^++\pi^-}$
    ($D_{c^{+}}^{\pi^++\pi^-}$) distribution at $\mu=5$~GeV. The upper
    panels display the absolute distributions while the lower ones
    their ratio to the central value of the baseline \texttt{MAPFF1.0}
    fit.}
  \label{fig:SIAonly}
\end{figure}

As mentioned in Sect.~\ref{sec:data}, we did not include the SLD
charm-tagged measurements because we have not been able to achieve an
acceptable description for this particular data set. Specifically, we
found that its inclusion causes a general deterioration of the fit
quality with a $\chi^2$ per data point of SLD-charm itself exceeding
6. We have identified the origin of this behaviour in an apparent
tension between the SLD charm-tagged and the COMPASS measurements. As
a matter of fact, if the COMPASS data is excluded from the fit, the
SLD charm-tagged data can be satisfactorily fitted. More precisely we
observe that the inclusion of COMPASS on top of SIA data leads to a
suppression of the
$D_{u^+}^{\pi^++\pi^-}=D_{u}^{\pi^+}+D_{\bar
  u}^{\pi^+}+D_{u}^{\pi^-}+D_{\bar u}^{\pi^-}$ distribution for
$z\gtrsim 0.1$ as compared to a fit to SIA data only. This behaviour
is visible in the left panel of Fig.~\ref{fig:SIAonly} where the
$D_{u^{+}}^{\pi^++\pi^-}$ distribution for the sum of positively and
negatively charged pions is displayed at $\mu=5$~GeV for the following
FF sets: the baseline \texttt{MAPFF1.0} fit (which includes SIA and
SIDIS data), a \texttt{MAPFF1.0}-like fit to SIA data only, and the
NNFF1.0 fit~\cite{Bertone:2017tyb} (which includes SIA data only). We
see that at intermediate values of $z$ the SIA-only {\tt MAPFF1.0} fit
and NNFF1.0 fit are in good agreement, while the baseline
\texttt{MAPFF1.0} fit is suppressed. As a consequence of this
suppression, the
$D_{c^{+}}^{\pi^++\pi^-}=D_{c}^{\pi^+}+D_{\bar
  c}^{\pi^+}+D_{c}^{\pi^-}+D_{\bar c}^{\pi^-}$ distribution of the
global \texttt{MAPFF1.0} fit gets enhanced to accommodate the
inclusive SIA data. This effect is visible in the right panel of
Fig.~\ref{fig:SIAonly} that shows for the $D_{c^{+}}^{\pi^++\pi^-}$
distribution a good agreement between the SIA-only fit and NNFF1.0
with the global \texttt{MAPFF1.0} fit being generally harder for
$z\gtrsim 0.1$.  This enhancement of the $D_{c^{+}}^{\pi^++\pi^-}$
distribution deteriorates the description of the SLD charm-tagged
data.  This is not surprising in that charm-tagged observables are
naturally sensitive to the charm FFs. We interpret the suppression of
$D_{u^{+}}^{\pi^++\pi^-}$ and the consequent enhancement of
$D_{c^{+}}^{\pi^++\pi^-}$ as an effect of the COMPASS data. We
conclude that the COMPASS and SLD charm-tagged data are in tension and
we decided to keep the former and drop the latter from our global fit.

We finally note that the suppression of the
$D_{u^{+}}^{\pi^++\pi^-}$ distribution also leads to a deterioration
of the description of the $uds$-tagged measurements from both DELPHI
and SLD that feature a $\chi^2$ per data point of respectively 2.84
and 2.05 (see Tab.~\ref{tab:chi2s}). However, this deterioration is
milder than that of the SLD charm-tagged data, thus we opted for
keeping the $uds$-tagged measurements in the fit.

\subsubsection{Impact of SIDIS data}
\label{sec:datavar}

\begin{figure}[!t]
  \begin{center}
    \includegraphics[width=0.9\textwidth]{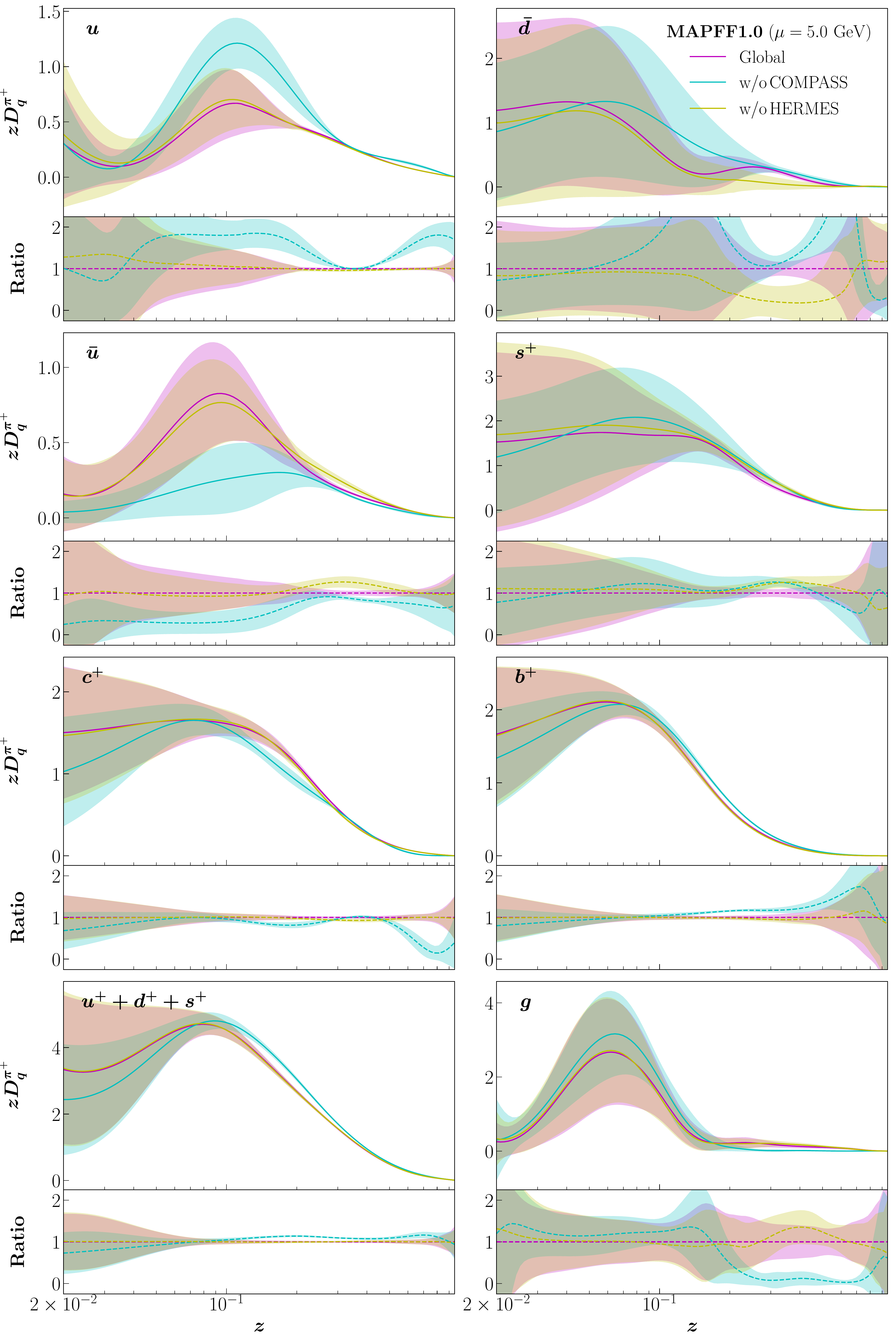}
  \end{center}
  \caption{\small Comparison between three variant of the
    \texttt{MAPFF1.0} fit: a fit to the global data set, a fit without
    the COMPASS data, and a fit without the HERMES data. The format of
    the plot is as in Fig.~\ref{fig:Baseline}.}
  \label{fig:DatasetsVariation}
\end{figure}

In this section, we study the impact of the individual SIDIS data sets
included in our analysis. To this purpose we have repeated our
baseline fit by removing either the COMPASS or the HERMES
measurements.  The three fits are compared in
Fig.~\ref{fig:DatasetsVariation}.

We note that the number of data points that survive the kinematic cuts
defined in Sect.~\ref{sec:data} is 8 for HERMES and 314 for COMPASS.
Despite the limited amount of data points, HERMES still provides a sizeable
constraint in the region of its coverage. As shown in
Fig.~\ref{fig:DatasetsVariation} the impact of the HERMES data in the region
$0.2 \lesssim z \lesssim 0.6$ can be summarised as follows:
\begin{itemize}
\item Overruling the COMPASS data for $D_{\bar{d}}^{\pi^+}$ and partly
  for $D_{\bar{u}}^{\pi^+}$. When excluding HERMES, these FFs are
  respectively suppressed by approximately 2-$\sigma$ and enhanced by
  1-$\sigma$. 
\item Competing with the COMPASS data for $D_{s^+}^{\pi^+}$ because
  both datasets have a comparable impact on this FF combination.
\item Overruled by the COMPASS data for the remaining FFs, namely
  $D_{u}^{\pi^+}$, $D_{c^+}^{\pi^+}$, $D_{b^+}^{\pi^+}$,
  $D_{g}^{\pi^+}$ and for the combination
  $D_{u^+ + d^+ + s^+}^{\pi^+}$, as their trend in the global fit
  follows that of the fit without HERMES.
\end{itemize}
We finally note that, as expected, the three fits display a similar
behaviour in the extrapolation regions for both the central value and
the uncertainty.

\subsubsection{Impact of SIDIS energy scale kinematic cut}
\label{sec:Qvar}

As discussed in Sect.~\ref{sec:data}, we included in our fit only
SIDIS data whose value of $Q$ is larger than $Q_{\rm cut}=2$~GeV. The
reason for excluding low-energy data stems from the fact that, as $Q$
decreases, higher-order perturbative corrections become increasingly
sizeable until eventually predictions based on NLO calculations become
unreliable. Therefore, $Q_{\rm cut}$ has to be such that NLO accuracy
provides an acceptable description of the data included in the fit.

\begin{figure}[!t]
  \begin{center}
    \includegraphics[width=0.65\textwidth]{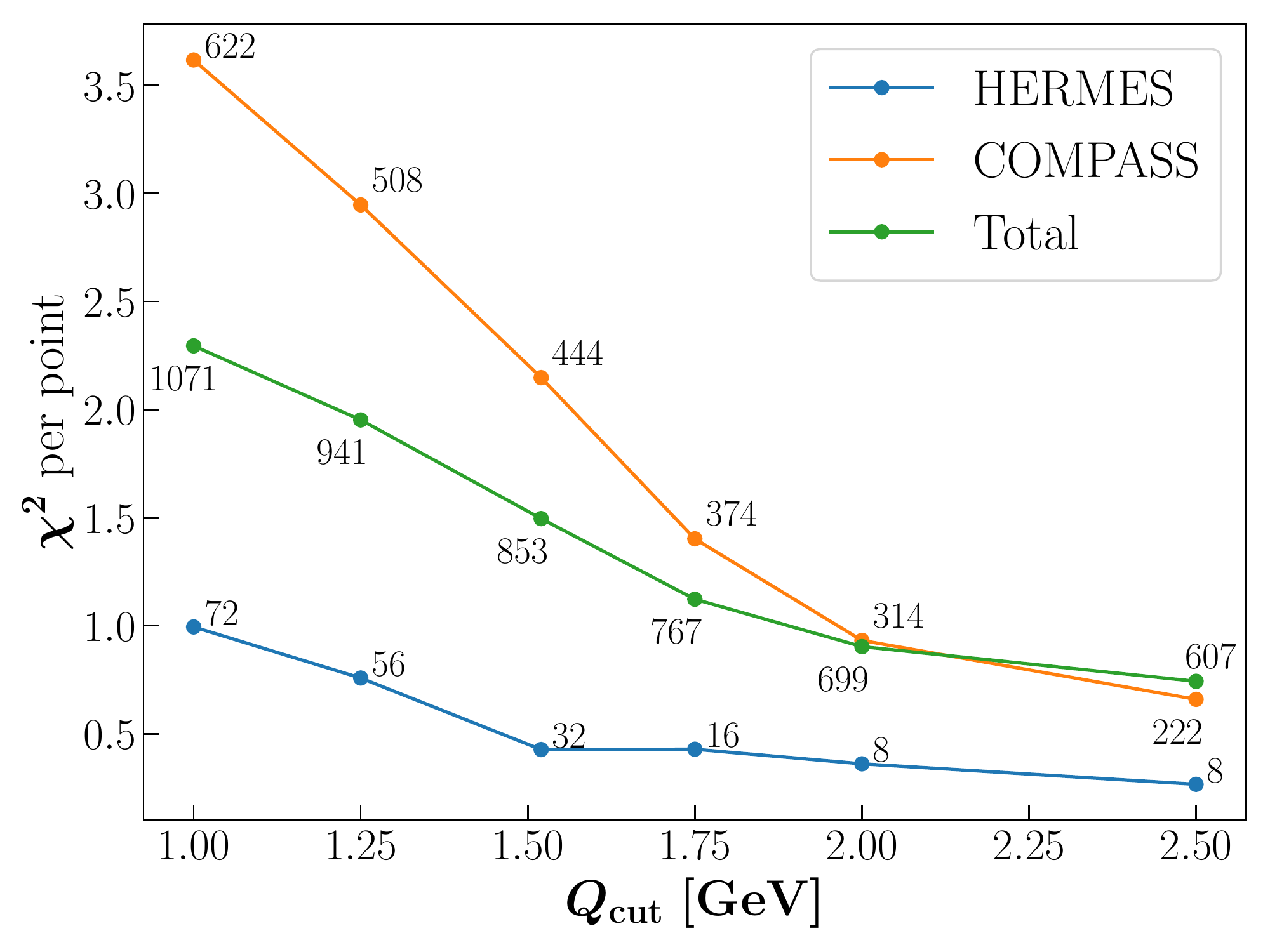}
    \end{center}
    \caption{\small Behaviour of the $\chi^2$ per data point as a
      function of the cut on $Q$, $Q_{\rm cut}$, applied to the SIDIS
      data. The $\chi^2$ is computed for the total \texttt{MAPFF1.0}
      data set (green curve), for the COMPASS data set (orange curve),
      and for the HERMES data set (blue curve). For each value of
      $Q_{\rm cut}$ considered, the plot also displays the number of
      data points $N_{\rm dat}$ that pass the cut.}
  \label{fig:SIDISQcut}
\end{figure}

Our particular choice is informed by studying the dependence of the
fit quality on the value of $Q_{\rm cut}$. To this purpose, we have
repeated our baseline analysis by varying the value of $Q_{\rm cut}$
in the range $[1.00,2.50]$~GeV. Fig.~\ref{fig:SIDISQcut} shows the
behaviour of the $\chi^2$ per data point for HERMES, COMPASS, and for
the total data set as functions of $Q_{\rm cut}$. For each point in
$Q_{\rm cut}$ the number of data points surviving the cut is also
displayed. As expected, the $\chi^2$ is a decreasing function of
$Q_{\rm cut}$ confirming the fact that perturbation theory works
better for larger values of the hard scale $Q$. However, while HERMES
can be satisfactorily described down to $Q_{\rm cut}=1$~GeV with a
$\chi^2$ that never exceeds one, the COMPASS $\chi^2$ quickly
deteriorates reaching a value as large as 3.5 at $Q_{\rm
  cut}=1$~GeV. Given the large size of the COMPASS data set, this
deterioration drives the total $\chi^2$ that also becomes
significantly worse as $Q_{\rm cut}$ decreases. Based on
Fig.~\ref{fig:SIDISQcut}, we have chosen $Q_{\rm cut}=2$~GeV for our
baseline fit because it guarantees an appropriate description not only
of the COMPASS data, but also of the entire data set.

\section{Summary and outlook}
\label{sec:conclusion}

In this paper we have presented a determination of the collinear FFs
of charged pions, dubbed {\tt MAPFF1.0}, based on a broad data set
that includes SIA and SIDIS data. Experimental uncertainties are
consistently treated by taking into proper account correlations
whenever available and propagated into the fitted FFs using the Monte
Carlo sampling method. Theoretical predictions are computed to NLO
accuracy in perturbative QCD and duly integrated over the relevant
final phase space when required in order to closely match the
experimental data. Appropriate kinematic cuts aimed at ensuring the
validity of the theoretical predictions for the fitted data have also
been enforced. Seven independent combinations of FFs are parametrised
in terms of a single neural network and fitted to data by means of a
trust-region algorithm making use of the knowledge of the analytic
derivatives of the neural network itself. The global $\chi^2$ as well
as the $\chi^2$ of all the single data sets included in the fits are
fully satisfactory. This is naturally reflected in a very good match
between experimental data included in the fit and the corresponding
theoretical predictions. We have compared the resulting {\tt MAPFF1.0}
FF set to two other recent determinations, {\tt DEHSS14} and {\tt
  JAM20}, finding some noticeable difference especially at the level
of the uncertainties.

In order to assess the stability and robustness of our results, we
have performed a number of variations w.r.t. the baseline settings.

As discussed in Sect.~\ref{sec:methodology}, we have explored
different numbers of independent FF combinations and selected the one
that implies a minimal set of restrictions in order to avoid any bias
deriving from too restrictive flavour assumptions. This resulted in a
set of 7 positive-definite independent combinations fitted to
data.

Given the dependence of the SIDIS predictions on PDFs, in
Sect.~\ref{sec:theochoices} we have discussed the effect on FFs of
including the PDF uncertainty as well as that of using different PDF
sets. As expected on the basis of the particular structure of the
SIDIS observable being fitted (multiplicities), we found that the
resulting FFs are almost insensitive to the treatment of PDFs both in
terms of uncertainties and central values.

In Sect.~\ref{sec:charmtag} we showed that the inclusion of SIDIS data
on top of the SIA data sets has the effect of ``rebalancing'' the
total up and total charm FFs with the consequence of substantially
worsening the description of SLD charm-tagged data. As a consequence,
the SLD charm-tagged data set has been excluded from the analysis.

In Sect.~\ref{sec:datavar} we have studied the interplay between SIA
and SIDIS data and observed that the latter, mostly represented by
COMPASS, plays a vital role in constraining and separating FFs
flavours. However, we also noticed that HERMES, despite the limited
number of points, has a noticeable impact on FFs.

Finally, in Sect.~\ref{sec:Qvar} we have justified our particular
choice for the cut on the minimum value of $Q$, $Q_{\rm cut}=2$~GeV,
for the SIDIS data included in the fit. We argued that this particular
value guarantees a reliable applicability of NLO accurate predictions
to the SIDIS data set. As a matter of fact, $Q_{\rm cut}=2$~GeV allows
us to obtain a global $\chi^2$ as well as the $\chi^2$ of the COMPASS
data set that are close to unity.\\

A possible natural continuation to this work is a determination of the
charged kaon, proton/antiproton and charged unidentified hadron
FFs. In all these cases, SIA and SIDIS data is available that would
allow for an extraction of these sets of distributions in a very
similar manner as done here for pions. In this respect, particularly
interesting is the measurement of the $K^-/K^+$ and $p/\overline{p}$
ratios recently presented by COMPASS~\cite{Akhunzyanov:2018ysf,
  Alexeev:2020jia}. These observables are affected by very small
systematic uncertainties and are thus promising to constrain kaon and
proton FFs.

In this work we have exploited the complementarity of SIA and SIDIS
observables to obtain and accurate determination and separation of the
quark FFs of the pion. However, both SIA and SIDIS are poorly
sensitive to the gluon FF because in both cases gluon-initiated
channels are only present starting from NLO. This is reflected in a
relatively large uncertainty of this distribution (see for example
Fig.~\ref{fig:Baseline}). In order to constrain the gluon FF, we plan
to use data for single-pion production in proton-proton collisions. As
proven in Ref.~\cite{Bertone:2018ecm}, this process is directly
sensitive to the gluon FF already at LO thus providing an effective
handle on this distribution.

Finally, we point out that an accurate determination of the
collinear FFs of the pions (as well as that of other light hadrons
such as the kaons and the protons) is instrumental to a reliable
determination of transverse-momentum-dependent (TMD)
distributions. Specifically, the description of
$p_{\rm T}$-dependent SIDIS multiplicities at low values of
$p_{\rm T}$, where $p_{\rm T}$ is the transverse momentum of the
outgoing hadron, can be expressed in terms of TMD PDFs and TMD FFs
that in turn depend on their collinear counterpart. The
HERMES~\cite{Airapetian:2012ki} and COMPASS~\cite{Adolph:2013stb}
experiments have measured this observable. This data can then be
used to extract TMD distributions extending, for example, the
analysis of TMD PDFs carried out in Ref.~\cite{Bacchetta:2019sam} to
the TMD FFs relying on the collinear FFs determined in this work.
This goal will be pursued by the MAP Collaboration in the future.
\\

The entirety of the results presented in this paper have been obtained
using the public code available from:
\begin{center}
{\tt\href{https://github.com/MapCollaboration/MontBlanc}{https://github.com/MapCollaboration/MontBlanc}.}
\end{center}
On this website it is possible to find some documentation concerning
the code usage as well as the FF sets in the {\tt LHAPDF} format. We
provide three sets of FFs with $N_{\rm rep}=200$ replicas each and of
their average for the positively and negatively charged pions and for
their sum. These are correspondingly called {\tt MAPFF10NLOPIp}, {\tt
  MAPFF10NLOPIm}, and {\tt MAPFF10NLOPIsum} and are made available
via the {\tt LHAPDF} interface~\cite{Buckley:2014ana} at:
\begin{center}
{\tt \href{https://lhapdf.hepforge.org/}{https://lhapdf.hepforge.org/}.}
\end{center}

\section*{Acknowledgments}
We thank W. Vogelsang for confirming that the misprint in the
$C_{L,qg}^{(1),nm}$ expression reported in Ref.~\cite{Stratmann:2001pb}
was corrected in Ref.~\cite{Anderle:2012rq}, and
Rodolfo Sassot for helping us debug the {\tt APFEL++}
implementation and for providing us with the predictions of
Ref.~\cite{deFlorian:2014xna}. We are grateful to Gunar Schnell
for support with the interpretation of the HERMES data and to Nobuo Sato and
the JAM collaboration for providing us with their theoretical predictions,
PDF and FF sets. We thank J. Rojo for a critical reading of the paper. The work of R.~A.~K.
is partially supported by the Netherlands Organization for Scientific
Research (NWO). V.~B. is supported by the European Union's Horizon
2020 research and innovation programme under grant agreement
\textnumero~824093. E.~R.~N. is supported by the UK STFC grant
ST/T000600/1 and was also supported by the European Commission through
the Marie Sk\l odowska-Curie Action ParDHonSFFs.TMDs (grant number
752748).

\bibliography{MAP10FF}

\end{document}